%% file: orbi.tex
\documentclass[12pt]{article}
\usepackage{latexsym}
\usepackage{amssymb}
\usepackage{amsfonts}
\usepackage{epsfig}
\usepackage{lscape}

\catcode`\@=11
\def\numberbysection{\@addtoreset{equation}{section}
        \def\theequation{\thesection.\arabic{equation}}}

\def\be{\begin{equation}}
\def\ee{\end{equation}}
\def\ba{\begin{eqnarray}}
\def\ea{\end{eqnarray}}

\def\sc{\scriptsize}

\def\ov{\overline}

\def\R{{\rm Re}\ }
\def\Z{\mathbb{Z}}
\newcommand{\nl}{\nonumber \\}

\def\u1{\widehat{U(1)}}
\def\su2{\widehat{SU(2)}}
\def\TT{{\bf T }}
\def\OO{{\bf O }}
\def\II{{\bf I }}
\def\CC{{\bf C }}
\def\DD{{\bf D }}

\def\ap{\a'}

\def\ra{\rangle}
\def\rra{\rangle\!\rangle}

\def\a{\alpha}
\def\b{\beta}
\def\g{\gamma}
\def\G{\Gamma}

\def\d{\delta}
\def\e{\epsilon}
\def\z{\zeta}
\def\h{\eta}
\def\th{\theta}

\def\l{\lambda}
\def\L{\Lambda}
\def\m{\mu}

\def\x{\xi}

\def\p{\pi}

\def\r{\rho}
\def\s{\sigma}

\def\t{\tau}
\def\f{\phi}

\def\c{\chi}
\def\w{\omega}
\def\W{\Omega}
\def\Th{\Theta}

\numberbysection

\begin{document}
\begin{titlepage}
\begin{center}
\hfill  \quad DFF 378/12/2001 \\
\hfill  \quad PAR-LPTHE 01-79 \\

\vspace{2.cm} 

{\Large \bf Boundary States of $c=1$ and $3/2$ } \\ 
\vspace{.3cm}
{\Large \bf Rational Conformal Field Theories} 

\vspace{1cm}

\qquad Andrea CAPPELLI${}^a$ and
Giuseppe D'APPOLLONIO${}^{b}$ \\
\vspace{.5cm}
{\em ${}^a$ I.N.F.N. and Dipartimento di Fisica}\\
{\em  Via G. Sansone 1, 50019 Sesto Fiorentino - Firenze, Italy} \\
\medskip
{\em ${}^b$ Lab. de Physique Th\'eorique et Hautes Energies}\\
{\em Universit\'e Pierre et Marie Curie, Paris VI}\\
{\em  4 Place Jussieu, \ 75252 Paris Cedex 05, France}

\end{center}

\vspace{.5cm}
\begin{abstract}
We study the boundary states for the rational points
in the moduli spaces of $c=1$ conformal and  
$c=3/2$ superconformal field theories, including 
the isolated Ginsparg points.
We use the orbifold and simple-current techniques to relate the
boundary states of different theories and 
to obtain symmetry-breaking, non-Cardy boundary states.
We show some interesting examples of fractional and twisted
branes on orbifold spaces.
\end{abstract}

\vfill
\end{titlepage}
\pagenumbering{arabic}


\section{Introduction}

One of the topics of present string theory investigations
is the determination of D-branes on general backgrounds: 
their algebraic construction as boundary states of
conformal field theory, their classification,
the relevant consistency conditions and their geometrical interpretation 
\cite{rs}\cite{as}\cite{douglas}\cite{mms}\cite{rev-bs}.

In the rational conformal theories (RCFT), the boundary states should
obey a closed set of modular covariance conditions 
\cite{cardy}\cite{comcon}\cite{zuber}, that
involve the bulk-theory data specified by the torus
partition function. A general solution was originally found by
Cardy for the symmetry-preserving boundary states associated to the
charge-conjugation partition function \cite{cardy};
several authors have recently discussed other cases and examples with
symmetry-breaking boundaries and general partition functions
\cite{su2}\cite{ca}\cite{sbb}\cite{gaber}.
Among other solutions, a rather interesting pattern has emerged for
the D-branes on group manifolds \cite{as} and coset manifolds 
\cite{mms}\cite{eli}.
Moreover, a theory for the symmetry-breaking boundary conditions 
has been introduced \cite{sbb} and a large class of non-Cardy boundary
states has been found \cite{bcs} by extending the method of simple currents
\cite{simple}.

In this paper, we would like to contribute to these searches by presenting
the detailed analysis of the rational conformal and superconformal
theories at $c=1$ and $3/2$, respectively \cite{gins}\cite{dgh}; 
our study includes the isolated points of the non-abelian orbifolds  
$SU(2)/G$, where $G$ is the symmetry group of the
tetrahedron ($\TT$), octahedron ($\OO$) and icosahedron ($\II$).
These theories are interesting for their non-trivial,
yet manageable, chiral algebras, involving several twisted sectors
\cite{dvvv}.
Their boundary states provide nice examples of 
``fractional'' and ``twisted'' branes \cite{fracbranes}.
In our analysis of boundary states, we extensively use the method 
of simple currents \cite{bcs}, and 
we exemplify some properties of symmetry-breaking boundaries 
first discussed in Ref.\cite{sbb}.

It is interesting to see these methods at work in
elaborate examples and to discuss the resulting features. Whenever
orbifold constructions, often implemented by simple currents,
map pairs of conformal theories, we can find 
corresponding relations between the respective boundary states. 
Cardy-type boundaries are mapped to new, non-Cardy boundaries 
pertaining to the same or a different theory; these relations provide
interesting hints and checks for the geometrical interpretation
of the D-branes.

The paper is organized as follows. 
In Section 2, we introduce the orbifold construction of boundary
states in the rather well-known case of the $\su2_k$ affine conformal
theories \cite{su2}\cite{comcon}\cite{zuber}, 
whose bulk field content is given by the $ADE$ modular invariants
\cite{ciz}; this example also motivates the
general formulae for the orbifold constructions based on simple currents
\cite{bcs}\cite{s-klein}.  
In Section 3, we discuss the boundary states for the rational $c=1$ theories
of the compactified boson and the $S^1/\Z_2$ orbifold \cite{gins}; most
of the results are well known\footnote{
See e.g. the Refs.\cite{rs}.}, 
but they set the stage for 
the analysis of the $\TT-\OO-\II$ models.

In Section 4, after (re)-deriving the $\TT-\OO-\II$
chiral algebras and $S$ matrices \cite{dvvv}, we
discuss an interesting example of non-Cardy boundary states 
that pertains to the tetrahedron orbifold with
diagonal partition function. Such boundaries are derived
from the Cardy states of the octahedron by simple-current extension,
namely by the inverse of the orbifold map:
$\su2_1/\OO =( \su2_1/\TT )/\Z_2$.
The corresponding annulus amplitudes provide a 5-dimensional
representation of the 21-dimensional fusion rules of the $\TT$ model.

In Section 5, we review the moduli space of $c=3/2$ superconformal 
theories \cite{dgh}, we write their chiral algebras and characters, 
and find their boundary states.
Finally, Section 6 is devoted to the superconformal 
$\TT$ and $\OO$ orbifolds\footnote{
The superconformal $\II$ orbifold is not discussed here.}; 
using the chiral data spelled out in Appendix \ref{stoi-ch}, 
we find the boundary states for some of their non-charge-conjugate
partition functions and the relations among them.

The Appendices contain some details of our work: Appendix A
reports the character tables of the $\TT-\OO-\II$ groups; 
Appendix B discusses the amplitudes for 
non-orientable surfaces, the Klein bottle and the 
M\"obius strip, \cite{carg} that complete the analysis of $c=1$ theories.
Appendix C and D contain the chiral algebras of 
the conformal and superconformal $\TT-\OO-\II$ models, respectively.


\section{Orbifold constructions for boundary states}

Rational conformal theories are characterized by
the modular invariant partition function on the torus,
that is a sesquilinear form in the characters $\chi_i$ of the 
representations of the chiral algebra 
${\cal A}$ of the theory \cite{rev-cft}:
\be
Z = \sum_{i,j=1}^N Z_{ij}\ \chi_i(q) \ov{\chi_j(q)} \  ,\qquad\qquad
q=\exp( 2i\pi \t)\ .
\label{modinv}
\ee
In this expression, the trace over the states propagating in the bulk
decomposes into the representations labeled by the indices $i$ and $j$
that occur with integer multiplicities $Z_{ij}$.

The determination of the conformal boundary conditions that are consistent
with a given, generic bulk theory (generic $Z$) 
is a non-trivial problem that has been tackled by 
the recent literature \cite{su2}\cite{comcon}\cite{ca}\cite{zuber}.
Let us briefly recall the setting:
the partition function on the annulus with boundary conditions
of type $(a)$ and $(b)$ is a linear combination of characters,
\be
A_{ab} =  \sum_{i=1}^N A^i_{ab}\ \chi_i(q) \ ,
\label{Aab}
\ee
where the non-negative integer multiplicities $A^i_{ab}$ are 
in general unknown, and should be determined by the consistency conditions of 
modular covariance and by some physical requirements \cite{cardy}.
Upon performing the $S$ modular transformation, $\t \to -1/\t$, 
the annulus amplitude (\ref{Aab}) describes the propagation
of bulk states between the two boundary states
$|a\ra$ and $|b\ra$. The latter can be expended in the
basis of the Ishibashi states $ |m \rra $ as follows \cite{ishi}:
\be
\vert a \ra = \sum_{m=1}^M B_{am}\ \vert m \rra \ .
\label{bcoeff}
\ee 
There exists an Ishibashi state for any bulk representation that
reflects at the boundary, namely for any $m$ 
such that $Z_{mm^*} \ne 0$ in the partition function (\ref{modinv}),
with corresponding multiplicity ($m^*$ is the 
representation conjugate to $m$).

The general form of the boundary states for the bulk 
theory of the charge-conjugation modular invariant, 
$Z_{ij} = \d_{i,j^*}$, has been given by Cardy \cite{cardy}:
there are as many boundary states as 
representations of the chiral algebra, $a,m=1,\dots, N$, and
the boundary coefficients are expressed in terms of the 
modular $S$ matrix,  
\be
B_{am} = \frac{S_{am}}{\sqrt{S_{0m}}} \ .
\label{cardy-b}
\ee
As a consequence of the Verlinde formula \cite{verlinde}, 
the corresponding annulus 
coefficients are equal to the fusion rules: $A^i_{ab}= {\cal N}^i_{ab}$.

For general torus partition functions,
we can make the natural assumption of completeness of the
boundary conditions \cite{comcon}, such that the boundary coefficients
$B_{am}$ define an invertible map in (\ref{bcoeff}).
Orthogonality is also required for the set of ``pure'' boundaries
whose correlators obey the cluster decomposition.
These conditions imply that the matrix
$R_{am} \equiv \sqrt{S_{0m}}\ B_{am}$ is unitary.
An equivalent condition is that the matrices of annulus 
coefficients\footnote{
Note that the bulk $(i)$ and boundary $(a,b)$
indices of $A^i_{ab}=A^i_{ba}$ can be raised
with the help of the bulk and boundary conjugation matrices,
$C_{ij}=(S^2)_{ij}=\d_{i,j*}$ and $A^0_{ab}=\d_{a,b^*}$, respectively.}
$(A_i)_a^{\ \ b}$ 
give rise to an integer-valued representation of 
the fusion algebra \cite{comcon,zuber}:
\be
\sum_{b=1}^{M}\ A^{\ \ b}_{ia}\ A^{\ \ c}_{jb} = 
\sum_{k=1}^{N}\ {\cal N}_{ij}^{\ \ k}\ A^{\ \ c}_{ka} \ .
\ee

In this paper, we study boundary states at $c=1$ and $3/2$: we discuss
interesting theories not completely analyzed so far, and describe cases with
non-charge-conjugate modular invariants.
We extend the orbifold constructions of bulk theories
to the determination of the boundary states,
relying on the results of the Refs.\cite{sbb,bcs,mms}. 
The method can be illustrated in the case of the $\su2_k$ 
affine conformal theories with $ADE$ modular invariant partition functions
\cite{ciz}.
The $D$-theories with non-diagonal modular invariant
can be obtained as orbifolds of the diagonal $A$-theories, by
modding out the $\Z_2$ symmetry $\chi_i \mapsto (-1)^{i+1} \chi_i$,
$i=1,\dots,k+1$. 
The invariant states are the integer-spin $\su2_k$ representations, and 
the twisted states are added in a way that is consistent with
modular invariance.
The result is ($\ell=1,2,\dots$):
\ba
\!\!\!\! k=4\ell+2,\ \ \ \ \ 
Z_{D_{2\ell+3}} &\!=\!& \sum_{i=1 \ odd}^{k+1} |\chi_i|^2 + 
|\chi_{(k+2)/2}|^2 + 
\sum_{i =2 \ even}^{(k-2)/2}\left(
\chi_{i}\ov{\chi}_{k+2-i} + {\rm c.c.}\right)\ ;
\nl
\!\!\! k=4\ell,\ \ \qquad\quad 
Z_{D_{2\ell+2}} &\!=\!& \sum_{i =1\ odd}^{(k-2)/2} 
| \chi_i + \chi_{k+2-i}|^2 + 2|\chi_{(k+2)/2}|^2 \ . 
\label{d-inv}
\ea
The $D$-even partition functions are diagonal modular invariants for
an extended chiral algebra, while the $D$-odd partition functions
have left and right sectors paired by the permutation $i \to k+2-i$,
that is an automorphism of the fusion rules.
For $k=6$ in particular, the orbifold construction
relates the $A_7$ and $D_5$ partition functions:
\be
Z_{A_7} = \sum_{i=1}^7\ \left| \c_i \right|^2\ \ \ \longrightarrow\ \ \ \
Z_{D_5} = \sum_{i=1 \ odd}^7\ \left| \c_i \right|^2 +
   \left| \c_4 \right|^2 +\left( \c_2 \ov{\c}_6 + {\rm c.c.} \right)\ .
\label{d5}
\ee
The even-index characters in $Z_{D_5}$ correspond to the twisted
sectors.

The orbifold operation can be applied to the Ishibashi states
of the diagonal invariant since they
are in one-to-one relation with the representations of the chiral algebra.
First we should form combinations of boundary states 
that are invariant under the orbifold symmetry;
in the present case, they are given by:
\be
|i \ra_D = \frac{1}{\sqrt{2}} \left ( |i \ra_A + |k+2-i \ra_A \right ) 
= \sum_{m=1}^{k+1}\ \frac{S_{im}}{\sqrt{S_{1m}}} 
\frac{1+(-1)^{m+1}}{\sqrt{2}} | m \rra\ ,\qquad 
i = 1,\dots, \frac{k}{2} \ ,
\label{inv-bs}
\ee
where $S_{ij}=\sqrt{2/(k+2)} \ \sin(\pi ij /(k+2))$.
The states $|i\ra_D$ 
can be called ``invariant'' boundaries because
they only allow the propagation of odd-$m$ bulk states in the closed channel.

In the $D$-odd theories, we need two further boundaries 
to form a complete basis.
These arise from the splitting of the boundary 
state of the diagonal theory $|f \ra_A$, $f=(k+2)/2$,  that 
is the fixed point of the orbifold action $i \to k+2 -i$:
\be
\left | f, \pm \right \ra_D = \frac{1}{\sqrt{2}} \left |f \right \ra_A \pm 
\frac{R}{\sqrt{S_{1 f }}}\left |f \right \ra \! \ra \ .
\label{fix-bs}
\ee
These two boundary states 
are distinguished from the invariant combinations
(\ref{inv-bs}) by having non-vanishing coefficients for the
Ishibashi $| f \rra$ corresponding to
the ``twisted'' sector $i=(k+2)/2$ in the $D$-odd modular invariants
(\ref{d-inv}).

The coefficient $R =1/\sqrt{2} $ is
determined by the completeness condition for these boundaries 
(\ref{inv-bs}, \ref{fix-bs}), 
i.e. by the orthogonality of the matrix $R_{am}$. 
The two boundary states $|f,\pm\ra_D$ can be called 
{\it fractional branes}, using a string terminology \cite{fracbranes}:
owing to the splitting,
their boundary coefficients are smaller than those of the invariant
state $\sqrt{2}\ |f\ra_A$ they originate from.
Actually, there is a geometric interpretation of the 
$\su2_k$ $A$-type boundary states as
Dirichlet two-branes (two-spheres) on the $SU(2)$ manifold, 
the $S_3$ sphere, \cite{as}; in this picture, the previous orbifold action 
(\ref{inv-bs}) is the antipodal map, and 
the $|f\ra_A$ brane, localized at the equator, is left invariant
and gets split.
Simpler examples of this phenomenon will be found later among the
$c=1$ theories.

In the $D$-even theory, there are two degenerate Ishibashi
states $|f,\pm \rra$ corresponding to the fixed point $f=(k+2)/2$
of the orbifold action. One can similarly construct the
invariant boundary states and  the two fractional states; however, 
their boundary coefficients are not completely
determined by the completeness condition.
There remains a free rotation in the space of the states $|f,\pm \rra$, 
that are degenerate in Virasoro dimension; a proper basis 
for these states is the one preserving the extended
symmetry of the $D$-even theory. Such basis can be found by applying
the Cardy formula (\ref{cardy-b}) to the boundary states of
the extended theory, involving the $S$-matrix for the extended 
characters \cite{fss}.
The boundary states for all the $ADE$ modular invariants have
already been obtained by several methods \cite{su2,ca,zuber}, 
and the present discussion was just meant to be pedagogical; 
note that the orbifold construction can also be extended to
the boundary operator-product expansion \cite{ope}.

The orbifold construction can be generalized using the language of
simple currents \cite{simple}.
A simple current $J$ is a primary field with one-term
fusion rules with all the fields:
\be
J \cdot \f_i = \f_{J(i)} \ , \qquad\qquad i=1,\dots, N.
\label{j-def} 
\ee
The presence of the simple current implies an abelian discrete symmetry
in the theory, that is generated by $\exp(2i\pi Q_J)$, with:
\be
Q_J\left(\f_i \right) = h_J + h_i - h_{J(i)} \qquad {\rm mod}\ 1 .
\label{q-def}
\ee
This charge is the exponent for the monodromy of the current around
the field $\f_i$ and is conserved in the fusion rules.
The fields $\f_i$ can be organized in orbits, each orbit containing the 
fields generated by the repeated fusion with the simple current.
The simple current and its powers generate an abelian
group by fusion that is called the {\it center} ${\cal G}$ of the 
conformal field theory.

Starting from the charge-conjugation partition function,
one can obtain a new modular invariant by
modding out the abelian symmetry associated to the simple current.
The result depends on the order of 
the center and on the (conformal) spin of the current. 
In particular, an integer-spin current $J$  generates an
extension-type modular invariant of the form \cite{fss}:
\be
Z = \sum_{{\rm orbits}\ a  |\ Q_J(a) = 0}\ \left|{\cal S}_a\right| \ 
\left | \sum_{J \in {\cal G}/{\cal S}_a} \chi_{J(i_a)} \right |^2 \ ;
\label{ext-z}
\ee
in this equation, $a$ labels the orbits, $ i_a$ 
is a representative point on each orbit, and  $|{\cal S}_a|$ is the order of
the stabilizer ${\cal S}_a$ of the orbit $a$, i.e. the subgroup
of ${\cal G}$ acting trivially on any element $i$ in $a$.
Extension-type modular invariants can be considered as diagonal
 invariants with respect to the basis of the extended chiral algebra,
and therefore the Cardy solution (\ref{cardy-b}) can be used to obtain
boundary states that preserve the extended symmetry.
However, in many cases it is interesting to know also the boundaries
that only respect the original chiral algebra \cite{sbb}.

Another example of simple-current modular invariant
is of the automorphism type \cite{simple}:
\be
Z = \sum_{i \ | Q_J(i) = 0} |\chi_i|^2 \ +  
\sum_{i \ | Q_J(i) = 1/2} \chi_i \ov{\chi}_{J(i)} \ ,
\label{auto-z}
\ee
and it is generated by an order-two current of half-integer spin. 
Both types of invariants, (\ref{ext-z}) and (\ref{auto-z})
are realized in the $\su2_k$ $D$-series seen before, the simple current
being the primary field $\f_{k+1}$.

The $\mathbb{Z}_2$ automorphism modular invariant (\ref{auto-z})
will appear frequently in our analysis of the boundary states  
of $c=1$ and $c=3/2$ theories. 
The corresponding boundary coefficients $R_{am}$
have been found in general \cite{ca}, and were shown to represent 
the so-called {\it classifying algebra} for boundary conditions that 
replaces the fusion algebra for the Cardy case.
The general pattern of the boundary states  
is already apparent in the $\su2_k$ example: 
there are $\Z_2$-invariant boundaries that are in one-to-one
correspondence with length-two orbits of the simple current, 
\be
| a \ra =\sum_i \frac{S_{a,i}+S_{J(a),i}}{\sqrt{2S_{0,i}}} | i \ra \! \ra \ .
\label{inv-b}
\ee  
These states are characterized by having vanishing coefficients 
for all the Ishibashi states $|i\rra$ with $Q_J(i) = 1/2$, 
owing to the relation $S_{J(a),k}=S_{a,k} \exp(2\pi i Q_J(k))$
\cite{simple}. 
In addition there are fractional boundary states, two for each 
fixed point of the simple current, $J(f)=f,J(g)=g,\dots$, of the form:
\be 
| f, \pm \ra = \sum_i\ \frac{R_{f_{\pm},i}}{\sqrt{S_{0i}}}\ | i \rra  \ .
\label{frac-b}
\ee 
These states are characterized by non-vanishing 
coefficients on the Ishibashi corresponding to fields with
$Q_J=1/2$ and fixed by $J$, that 
can be expressed in terms of a suitable fixed-point $S$ matrix 
denoted by $\widetilde{S}$ \cite{bcs}:
\be
R_{f_{\pm},g} =  \pm \frac{\widetilde{S}_{fg}}{\sqrt{2}} \ .
\label{fix-s}
\ee
The coefficients in (\ref{frac-b}) with respect to 
the Ishibashi states $ | i \ra \! \ra $ with
$J(i) \ne i$ are simply given by the $S$ matrix of the model,
$R_{f_{\pm}i} =S_{fi}/\sqrt{2}$.
This ansatz for the boundary states was shown to
satisfy the previous constraints of integrality and positivity 
for the annulus amplitude \cite{ca}. We finally mention that
a general formula has been presented in Ref.\cite{bcs} for the boundary states
of arbitrary simple-current modular invariants.


\section{Boundary states at $c=1$: circle and orbifold lines}

We now turn to the analysis of the 
boundary states for the conformal field theories at $c=1$:
we should first recall some basic facts about these theories \cite{gins}. 
The first line of $c=1$ models is realized by the
free boson field $X$ compactified on a circle of radius $R$. These models
possess the affine $U(1)$ symmetry and
their field content can be organized in representations of this 
algebra, as summarized by the partition function:
\be
Z_c(R) = \sum_{n, m \in \Z} \ \G_{n,m}\ ,\qquad\qquad
\G_{n,m} = \frac{1}{|\eta (q)|^2 }  \ 
q^{\frac{\ap}{4} \left( \frac{n}{R}+\frac{mR}{\ap}  \right)^2} \ 
\ov{q}^{\frac{\ap}{4}\left( \frac{n}{R}-\frac{mR}{\ap}  \right)^2} \ ,
\label{z-circ}
\ee
where $\eta$ is the Dedekind function. 
By modding out the circle by the reflection ${\cal P}: X \mapsto -X$ 
one obtains the second line of the $S^1/\Z_2$ orbifold theories.
On each line, the points $R$ and $\ap/R$ are equivalent by T-duality
and the two lines intersect at one point, corresponding to 
 $R_c^2=4 \ap$ and $R_o^2=\ap$.
The circle theory at the self-dual 
radius $R^2_c=\ap$ possesses an $\widehat{SU(2)}_1$ 
affine symmetry that can be modded
by the discrete subgroups of $SU(2)$. 
While the orbifolds by the cyclic and dihedral groups reproduce theories 
on the circle and orbifold lines, respectively, 
those by the three non-abelian groups $\TT$, $\OO$, $\II$, 
(the symmetry groups of the tetrahedron, 
octahedron (cube) and icosahedron (dodecahedron)), 
give three new $c=1$ CFT, that do not belong to the previous lines
and have no marginal deformations.
In the sequel, we will focus on the rational points on the $c=1$ 
moduli space\footnote{
For recent results on the boundary states preserving the Virasoro algebra
only, see Ref.\cite{gaber}.} \cite{kiritsis}.


\subsection{Circle line}

There are two natural boundary conditions $\ov{J}=\W(J)$
for the $\u1$ currents $J$ and $\ov{J}$, namely  $\W(J)=\pm J$,
corresponding to Neumann and Dirichlet conditions, respectively. 
The corresponding Ishibashi states are:
\ba
|n,0 \ra \! \ra_D &=& \exp\left(\sum_{j=1}^{\infty} 
\frac{\a_{-j}\ov{\a}_{-j}}{j}\right) |n,0 \ra  \ , 
\nl
|0,m \ra \! \ra_N &=& \exp\left(-\sum_{j=1}^{\infty} 
\frac{\a_{-j}\ov{\a}_{-j}}{j}\right) |0,m \ra \ ,
\label{DN-ishi}
\ea
where $|n,m \ra$ is the highest weight state with $n$ units of momentum
and $m$ units of winding and the $\a_n$, $\bar{\a}_n$ are the bosonic modes. 
The boundary states  \cite{rs}\cite{rev-bs},
\ba
|x \rangle &=& \left ( \sqrt{\frac{\ap}{2}}\frac{1}{R} \right )^{1/2} 
\sum_{n \in \Z}\ e^{\frac{inx}{R}} \
|n,0 \ra \! \rangle_D \ , \nl
|\theta \rangle &=& \left ( \frac{R}{\sqrt{2 \ap}} \right )^{1/2} 
\sum_{m \in \Z}\ e^{im \theta} \ 
|0,m \ra \! \rangle_N \ ,
\label{DN-bound}
\ea
can be interpreted
as D0 and D1 branes, respectively; they belong to two continuous 
families parametrized by the position on the circle $x$ 
and by the value of a Wilson line $\th$.
 
Let us now focus on the RCFT at the radii
$R=\sqrt{\ap k}$, $k \in \mathbb{N}$, where the $\u1$ chiral algebra is
extended by the fields $e^{\pm i 2\sqrt{k/\ap}X}$ and
is usually referred to as the $\u1_k$ algebra \cite{rev-cft}.  
There are $2k$ primary fields, whose characters and conformal
dimensions are given by:
\be
\chi_r = \frac{1}{\eta (q)}\sum_{n \in \Z} q^{k(n-\frac{r}{2k})^2} \ , 
\hspace{1cm} r=-k+1,...,k \ ,
\hspace{1cm} h_r = \frac{r^2}{4k}\ . 
\label{u1-char}
\ee

The theory with charge conjugation modular invariant 
$Z=\sum_r \chi_r \ov{\chi}_{-r}$
possesses $2k$ boundary states that are specialization of the 
Dirichlet boundaries (\ref{DN-bound}) for $x= 2 \pi R(r/2k)$. 
In order to account for Neumann states in the rational theories,
we should consider symmetry breaking boundary conditions: 
we postpone this discussion to the next section.
The bulk theory described by the diagonal partition function 
$Z = \sum_r |\chi_r|^2$ can only have two boundary states, 
that are written as follows,
\be
|\pm \ra = \left ( \frac{k}{2} \right )^{1/4} 
\left ( |0 \ra \! \ra \pm |k \ra \! \ra \right )    \  ,
\label{diag-bound}
\ee
in terms of the Ishibashi states corresponding to the two self-conjugate 
fields, $r= 0, k$, of the rational theory. 
They can be realized as superpositions of D0 branes of the type 
(\ref{DN-bound}), sitting at even (resp. odd) multiples of  
$2 \pi R /(2k)$ \cite{mms}. 
This is the first occurrence of the orbifold relations of the previous
Section: actually, one can obtain the 
diagonal modular invariant from the charge conjugation one by the 
$S^1/ \Z_{k}$ orbifold of the symmetry
$\chi_l \mapsto \exp{( 2 i \pi l/k)} \chi_{l}$; thus,
the new boundary states are given by invariant combinations of the old ones,
in agreement with Eq. (\ref{inv-bs}).


\subsection{Orbifold line}

Let us first recall some general aspects of orbifold constructions 
\cite{dvvv} that will be useful in the following discussion.
An orbifold theory is the quotient ${\cal C}/G$
of the CFT ${\cal C}$ by a discrete group $G$ of 
symmetries of the theory, i. e.  by
an endomorphism group of the operator algebra, that
commutes with Virasoro and respects
the left-right decomposition of the Hilbert space:
\be
{\cal H} = \sum_{j,\ov{j}} [\f_j] \otimes [\ov{\f}_{\ov{j}}] \ ,
\label{h-spit}
\ee
where the $[\f_j]$ are the irreducible representations of the 
chiral algebra.
When a $\s$-model description is available, the target space
of the orbifold theory is the quotient of the original manifold 
by a subgroup $G$ of its isometry group.

The partition function of the ${\cal C}/G$ orbifold is: 
\be
Z = \frac{1}{|G|} \sum_{g,h \in G\ |\ [g,h]=0}\ 
Z \left [^g_h \right ]\ \e (g,h) \ , 
\label{orb-def}
\ee
where $|G|$ is the order of the group and $ Z \left [^g_h \right ]$  
is the contribution to the partition function coming from the
trace on the $g$-twisted sector with an insertion of the operator $h$. 
The restriction to commuting elements $ghg^{-1}h^{-1}=1$ comes from 
the fact that the cycle $aba^{-1}b^{-1}$ is contractible on the torus. 
The phase $ \e (g,h)$ accounts for the so-called
discrete torsion \cite{vafa}. 
There is a clear Hamiltonian interpretation: for each $g \in G$ 
there is an Hilbert space ${\cal H}_g$ of $g$-twisted states, that
are projected by 
$S_g = \{ h \in G | ghg^{-1}h^{-1}=1 \}$, the stabilizer of $g$.
Recalling the relation  $|G| = |S_g||C_g|$ between the order of the
stabilizer and the dimension of the conjugacy class of $g$,
we can restrict the sum in (\ref{orb-def})
to the conjugacy classes $a$ and choose a representative element $g_a$
for each class, leading to: 
\be
Z = \sum_{a} \frac{1}{|S_a|} \sum_{h \in S_a} Z[^{g_a}_h] \e (g_a,h)= 
\sum_a Tr_{{\cal H}_a} \Pi^a 
q^{L_0-\frac{c}{24}} \ov{q}^{\ov{L}_0-\frac{c}{24}} \ ,
\label{orb-def2}
\ee
where,
\be
\Pi^a =  \frac{1}{|S_a|} \sum_{h \in S_a} h \ \e (g_a,h)\ ,
\label{orb-proj}
\ee
is a projection operator onto $S_a$-invariant states.

The expression (\ref{orb-def}) of the partition function
does not show explicitly the chiral operator content of the orbifold theory,
that is necessary for the construction of the boundary states.
As stressed in Ref.\cite{dvvv}, several properties of the
orbifold chiral sectors can be associated to the representations 
of the finite group $G$.
With respect to its action, the original 
chiral algebra ${\cal A}$ decomposes 
into  ${\cal A}= \bigoplus_\a{\cal A_\a} $,
where ${\cal A_\a} $ contains the states that transform 
in the irreducible representation $r_\a$ of $G$. 
The chiral algebra of the orbifold is ${\cal A}_0={\cal A}/G$. 
Each  ${\cal A_\a} $ is a representation of
 ${\cal A}_0$, in general reducible because $G$ acts in  
${\cal A}_\a$ and commutes with  ${\cal A}_0$. 
We can then decompose each ${\cal A}_\a$ according to: 
\be
{\cal A}_\a = [\f_\a] \otimes r_\a \ ,
\label{orb-sect}
\ee
where $[\f_\a] $ is an irreducible representation of ${\cal A}_0$.

Representations of the chiral algebra that are mapped by  
$G$ into different representations are identified in the orbifold model; 
representations that are fixed point of $G$
are split and give rise to the twisted sectors ${\cal A}^g$,
that are in one-to-one correspondence with the conjugacy classes of $G$.
On ${\cal A}^g$,  it is defined the action of the stabilizer  
$S_g$,  and thus there is a decomposition analogous to (\ref{orb-sect}):
\be 
{\cal A}^g= \bigoplus_\a{\cal A}^{g}_\a \ , 
\hspace{1cm} {\cal A}^g_\a = [\f^g_\a] \otimes r^g_\a \ ,
\label{tw-sect}
\ee
where now $\a$ labels the irreducible representations $r^g_\a$ of $S_g$. 

The characters of the orbifold theory associated to the decomposition 
${\cal H}_g = \bigoplus_{\a} [\f_{\a}^g] \otimes r_{\a}^g$
can be written as combinations of the traces,
\be
z \left [^g_h \right ] = Tr_{{\cal H}_g} h\ q^{L_0-\frac{c}{24}} \ ,  
\hspace{1cm} 
\left |z \left [^g_h \right ] \right |^2 = Z \left [^g_h \right ] \ , 
\label{small-z}
\ee
as follows:
\be
\chi_{\a}^g(q) = \frac{1}{|S_g|} \sum_{h \in S_g} \r_{\a}^g(h^{-1})  
\ z \left [^g_h \right ]  \ ,
\label{chi-orb}
\ee
where $\r_{\a}^g(h^{-1})$ are the characters of the representation $r^g_\a$.
The characters  $\chi_{\a}^g(q)$ have $q$-expansion 
with positive integers coefficients. The inverse relation is: 
\be
z \left [^g_h \right ] = \sum_{\a}  \r_{\a}^g(h) \chi_{\a}^g(q) \ .
\ee 

The chiral algebra of the  $S^1/\Z_2$ orbifold at rational radius 
can be obtained using the previous formulas \cite{dvvv};
it contains $(k+7)$ primary fields whose characters read:
\be
\begin{array}{llll}
u_{\pm} &
= \frac{1}{2} \left ( \chi_0 \pm \sqrt{\frac{2 \h}{\th_2}} \right ) \ , 
&& h = 0,1 \ ,
\\ 
\f_{\pm} &= \frac{1}{2} \chi_k \ , && h = \frac{k}{4} \ ,
\\
\chi_r\ ,&  & \  r=1,..,k-1\ ,  & h = \frac{r^2}{4k} \ ,
\\
\s_i &= \frac{1}{2} \left( \sqrt{\frac{\h}{\th_4}} +  
\sqrt{\frac{\h}{\th_3}} \right) \ ,&\ i=0,1, & h = \frac{1}{16} \ ,
\\
\t_i &= \frac{1}{2} \left( \sqrt{\frac{\h}{\th_4}} -  
\sqrt{\frac{\h}{\th_3}} \right) \ ,&\ i=0,1, & h = \frac{9}{16}\ ,
\end{array}
\label{z2-chi}
\ee
where the characters $\chi_r$ are  defined in (\ref{u1-char})
and $\theta_\a$, $\a=2,3,4,$ are the Jacobi theta functions.
The fields\footnote{
Hereafter, the fields and the corresponding characters are labeled by
the same symbols.}
 $\chi_r$, $r=1,...,k-1$, arise from the identification 
between the primaries $\chi_r$ and $\chi_{-r}$. 
The fields $u_{\pm}$, $\f_{\pm}$ arise from the splitting of 
the representations at the two fixed points $r=0,k$: 
\ba
u_{\pm} &=&  Tr_{\chi_0} \left[
{1 \pm {\cal P} \over 2}q^{L_0-\frac{c}{24}} \right]
 \ ,  \nl
\f_{\pm} &=&  Tr_{\chi_k} \left[
{1 \pm {\cal P}\over 2}q^{L_0-\frac{c}{24}} \right] \ ,
\label{z2-split}
\ea
where the signs $\pm$ label the two irreducible representations of  $\Z_2$.
Finally there are four twisted fields $\s_i,\t_i$, $i=0,1$,
for the two fixed points, 
which are obtained by considering for each point the combinations:   
\be
\frac{1}{2} \left ( z \left [^{\cal P}_1 \right ] \pm 
z \left [^{\cal P}_{\cal P} \right ]  \right ) \ ,
\label{z2-twist}
\ee
in agreement with (\ref{chi-orb}).
The $S$ matrix in the basis (\ref{z2-chi}) was found in Ref.\cite{dvvv}:
for $k$ even, it reads (up to the factor $1/\sqrt{8k}$),
\be
\begin{array}{c|ccccc}
 & u_\pm & \f_\pm & \c_s & \s_j & \t_j
\\ \hline 
u_\pm & 1 &  1 & 2 & \pm\sqrt{k} & \pm\sqrt{k}
\\
\f_\pm & 1 & 1 & 2(-1)^s & \pm(-1)^j \sqrt{k} & \pm(-1)^j\sqrt{k}
\\
\c_r & 2 & 2(-1)^r & 4\cos(\pi r s/k) & 0 & 0
\\
\s_i& \pm\sqrt{k}& \pm(-1)^i\sqrt{k} & 0 & \d_{ij}\sqrt{2k}& -\d_{ij}\sqrt{2k}
\\
\t_i &\pm\sqrt{k} &\pm(-1)^i\sqrt{k} &0 & -\d_{ij}\sqrt{2k}&\d_{ij}\sqrt{2k}
\end{array}
\label{z2-s}
\ee
where $i,j=0,1$. The expression for $k$ odd is \cite{odd-orb}: 
\be
\begin{array}{c|ccccc}
 & u_\pm & \f_\pm & \c_s & \s_j & \t_j
\\ \hline 
u_\pm & 1 &  1 & 2 & \pm\sqrt{k} & \pm\sqrt{k}
\\
\f_\pm & 1 & -1 & 2(-1)^s & \pm i(-1)^j \sqrt{k} & \pm i(-1)^j\sqrt{k}
\\
\c_r & 2 & 2(-1)^r & 4\cos(\pi r s/k) & 0 & 0
\\
\s_i& \pm\sqrt{k}& \pm i(-1)^i\sqrt{k} & 0 & 
e^{i \s_{ij}\pi/8}\sqrt{k}& 
-e^{i \s_{ij}\pi/8}\sqrt{k} 
\\
\t_i &\pm\sqrt{k} &\pm i(-1)^i\sqrt{k} &0 & 
-e^{i \s_{ij}\pi/8}\sqrt{k}&
e^{i \s_{ij}\pi/8}\sqrt{k}
\end{array}
\label{zodd-s}
\ee
where $\s_{ij}=(-1)^{i+j}(-1)^{(k+1)/2}$, $i,j=0,1$.

The Cardy boundary states of the $S^1/\Z_2$ 
orbifold can be read from these $S$ matrices, Eqs. (\ref{bcoeff}) and
(\ref{cardy-b}); actually, they can be interpreted as the result of
the orbifold action on the boundaries of the 
circle theory \cite{rs,sbb}, in agreement with the discussion of 
Section 2.
The states $|\chi_r \ra$, $r=1,...,k-1$
come from the identification of D0 branes sitting at opposite points 
along the circle: 
in their spectrum there is an exactly marginal operator that allows 
displacements of the branes along the circle. 
The states $|u_+ \ra$, $|u_- \ra$, $|\f_+ \ra$ and $|\f_- \ra$ 
describe fractional branes D0 sitting at the fixed points of
the interval. 
Actually, they have smaller coefficients than those of the states
$|\chi_r\ra$ and non-vanishing coefficients for the twisted Ishibashi's;
these branes are forced to live on the fixed points  because
they have no marginal deformations,
unless they combine with the other fractional brane.
Finally the Cardy states corresponding to the twisted fields 
$|\s_i \ra$, $ |\t_i \ra$, $i=0,1$,
can be interpreted as fractional D1-branes with suitable Wilson lines.
Actually, the occurrence of both D0 and D1 branes is
not unexpected, because both gluing conditions $\W$ and $\W{\cal P}$ 
should be present in the orbifold theory.

A general feature of abelian orbifolds is that they always contain
a set of integer-spin simple currents, stemming from the decomposition of 
the chiral algebra of the original theory, 
that form a group isomorphic to the orbifold group $G$ \cite{simple}. 
These currents allow for reconstructing  the original theory as 
a simple current extension of the orbifold (see Eq.(\ref{ext-z})).
Among the chiral fields of the $S^1/\Z_2$ orbifold (\ref{z2-chi}), 
there is indeed the integer-spin current $u_-$ that gives back 
the circle theory.

According to the discussion of Section 2, we can use the simple-current
map to transform the orbifold boundary states back to the circle theory; 
according to Eq. (\ref{inv-b}), $|u_+ \ra_o$ and $|u_- \ra_o$ combine into
$| \chi_0 \ra_c$, while $|\f_+ \ra_o$ and $|\f_- \ra_o$ give 
$|\chi_k \ra_c$; the fixed points $|\chi_r \ra_o$ split 
giving the two boundary states $|\chi_r \ra_c$ and $|\chi_{-r} \ra_c$,
using Eq.(\ref{frac-b}).
These are the symmetry-preserving boundaries of the circle theory
seen before.

Furthermore, two other boundary states are obtained
from the fractional D1 branes $|\s_i\ra_o, |\t_i\ra_o$: 
\be
|+ \ra = \frac{1}{\sqrt{2}} ( |\s_0 \ra_o + |\t_0 \ra_o ) = 
\left( \frac{k}{2} \right)^{1/4} ( |u_+ \rra -  |u_- \rra + 
|\f_+ \rra -  |\f_- \rra )_o \ , 
\label{z2-diag}
\ee
and similarly for $|- \ra = ( |\s_1 \ra_o + |\t_1 \ra_o )/\sqrt{2}$.
The states $|\pm\ra$ preserve only  a $U(1)_k/\Z_2$ 
orbifold subalgebra of the full $U(1)_k$ symmetry of the circle \cite{sbb}
and can be interpreted as D1 branes with a particular Wilson line, 
namely they correspond to Neumann boundary conditions for the circle theory.  
This interpretation is confirmed by the expression of the
annulus amplitude between these new states 
and the rational circle D0 states $|\chi_r\ra$:
\be
A_{+, r} = \s_0 + \t_0  \ ,
\ee
which is the usual partition function for a free boson on a 
strip with Neumann-Dirichlet boundary conditions \cite{rev-bs}.
Actually the Ishibashi states  $|u_+ \rra -  |u_- \rra$ and 
$|\f_+ \rra -  |\f_- \rra$ are precisely 
those associated to the Neumann gluing condition $ \W {\cal P}$. 
According to the general analysis of symmetry breaking boundary
conditions \cite{sbb},  we can move the automorphism $r \mapsto -r$ 
from the gluing condition to the modular invariant
and consider the boundary states (\ref{z2-diag}) as the complete
set pertaining to the diagonal modular invariant of the circle theory.
Actually, they coincide with the boundaries (\ref{diag-bound}).

Two further simple currents exists in the orbifold theory, 
 $\f_{\pm}$, that, together with $u_-$, generate 
the $\Z_2 \times \Z_2$ group for $k$ even and the $\Z_4$ group for $k$ odd. 
Let us discuss the other theories that they may yield. 
We note that there exists another orbifold modular 
invariant given by the automorphism 
$(\f_+,\s_0,\t_0) \leftrightarrow (\f_-,\s_1,\t_1)$, which 
coincides with charge conjugation for $k$ odd.
The boundary states for this modular invariant at $R^2=\ap k$ 
can be constructed as before, by interpreting them as
symmetry-breaking boundaries for the original theory, i.e.
by shifting the automorphism from the modular invariant
to the gluing condition.
Consider the extension of the orbifold theory at $R^2 =4 \ap k$
given by the simple current $\f_+$ \cite{sbb}: 
the boundaries with $Q_{\f_+}=0$ at $R^2 =4 \ap k$ 
give back the $(k+7)$ symmetry-preserving Cardy boundaries 
of the theory at $R^2=\ap k$: 
these are obtained from the $k+1$ length-two orbits corresponding to
$(u_+,\f_+)$, $(u_-,\f_-)$ and $(\chi_r,\chi_{4k-r})$, with
$r$ even, and from the three fixed points $\chi_{2k}$, $\s_0$ and 
$\t_0$. Furthermore, the boundaries 
with  $Q_{\f_+}=1/2$ give the $k+1$ symmetry breaking boundaries
at $R^2=\ap k$ that we were after:
these are the $k$ length-two orbits 
$(\chi_r,\chi_{4k-r})$, with $r$ odd, and $(\s_1,\t_1)$.  
They can be interpreted as $k$ D0-branes at
$x_r = \pi rR/2k $, $r$ odd,  and as another D1-brane (the symmetry
preserving D0-branes are instead at $x_r = \pi rR/2k$, $r$ even). 
The explicit expressions of the symmetry-breaking boundaries can be read
from the matrices (\ref{z2-s}) and (\ref{zodd-s}); the boundaries
associated to the orbit $(\chi_r,\chi_{4k-r})$ with $r$ odd are,
for instance ($r=1,3,\dots,2k-1$):
\be
|\chi_r \ra = \frac{2}{(8k)^{1/4}}  \left (
(|u_+ \rra - |\f_+ \rra) +(|u_- \rra - |\f_- \rra) + 
\sqrt{2} \sum_{s=1}^{k-1} \cos \left (\frac{\pi rs}{2k} \right )
(|\chi_{2s} \rra - |\chi_{4k-2s}\rra) \right ) \ ,
\ee
where the Ishibashi states are those of the orbifold theory at $R^2=4\ap k$.
As said, these are also the boundaries of the
automorphism modular invariant.
Note that for $k$ odd, these boundaries can also be obtained
using directly the simple current $\f_+$ of the theory at $R^2 = \ap k$. 
Finally, for $k=4l+2$ the current $\f_+$ generates
another automorphism modular invariant given by the exchange
$\chi_r \leftrightarrow \chi_{k-r}$ for $r$ odd.

In conclusion, we have seen that the orbifold map and its
simple-current inverse relating the bulk theories
have a clear extension to the boundary states.
These maps allow to determine complete sets of boundaries for
non-charge-conjugate modular invariants and some
symmetry breaking boundaries; furthermore, they can indicate
a geometric interpretation of the boundary states in the orbifold theories.

The description of the circle and $S^1/\Z_2$ orbifold theories
can be completed by determining the appropriate
Klein and M\"obius amplitudes, that 
project the bulk and boundary spectrum under the action of the 
worldsheet parity operation \cite{carg}. 
These are briefly discussed in Appendix \ref{klein}.


\section{Boundary states of $c=1$ isolated points}


\subsection{Chiral algebras of $\TT-\OO-\II$ orbifolds}

Let us now consider the $\su2_1/G$ orbifolds of the circle theory
at the self-dual point, where $G$ is a discrete subgroup of
$SU(2)$ \cite{gins}.
The series of the cyclic groups 
$G=\CC_n$ have elements $g_{l/n}$, 
$l=0, ..., n-1$, that rotate of the angle 
$2 \pi l/n$ around the $J^3$ axis, 
where $J^i$, $i=1,2,3$, are the three $SU(2)$ generators.
On the boson field $X$, this action simply amounts to the shift
$X \mapsto X + 2 \p \sqrt{\ap} l/n$. 
The orbifold partition function is:
\be
Z_n \equiv Z(\CC_n) = \frac{1}{n} 
\sum_{k,l=0}^{n-1} Z\left [^{g_{k/n}}_{g_{l/n}} \right ]  \ ,
\ee
where
\be
Z\left [^{g_{k/n}}_{g_{l/n}} \right ]  
=  \frac{1}{|\eta (q)|^2} \sum_{p \in \Z , 
\ m \in \Z+\frac{k}{n}} e^{\frac{2 \p i l p}{n}}
q^{\frac{(p+m)^2}{4}}\ov{q}^{\frac{(p-m)^2}{4}} \ .
\ee 
Actually these orbifolds coincide (up to $T$ duality) with the 
compactified boson theories at the points $R^2 = \ap n^2$. 
The second series of orbifolds by the
dihedral groups $\DD_n$ similarly give points along the orbifold line at 
radius  $R^2 = \ap n^2$; actually, the $\DD_n$ groups 
are generated by adding the element 
$\exp(i \pi J_1)$ to $\CC_n$ whose action on the bosonic 
field is precisely the reflection $X \mapsto -X$. 

Finally there are the orbifold by the symmetry groups of the regular solids
$\TT,\OO,\II$, respectively $A_4$, $S_4$ and $A_5$, or
more precisely their lifts to $SU(2)$: 
$SL_2(\Z_3)$, $GL_2(\Z_3)$  and $SL_2(\Z_5)$
(In Appendix \ref{toi-tab} we report their character tables).
Following Ref.\cite{gins}, it is convenient to express the 
partition functions as sums over the abelian orbifolds
of the mutually commuting subgroups of the non-abelian groups,
with all overlappings removed.
The mutually commuting elements of the non-abelian
 groups $A_4$, $S_4$ and $A_5$, can be easily visualized in terms of their 
action on the tetrahedron, cube and dodecahedron, respectively. 
For the tetrahedral group we have $4$ $\CC_3$ subgroups that acts 
by rotations around axes through the centers of the faces and a 
$\DD_2$ generated by rotations of $\pi$ around axes passing through 
the center of opposite edges. The partition function is then:
\be
Z(\TT) = \frac{1}{12} \left ( 4 \left ( 3 Z_3 - Z_1 \right ) + 
4 Z_{o}(2\sqrt{\ap}) \right ) \ .
\label{z-tet}
\ee
A similar analysis for the octahedron and the icosahedron yields
the partition functions \cite{gins}:
\ba
Z(\OO) &=& \frac{1}{24} \left ( 3(4Z_4 - 2Z_2) + 4(3Z_3 - Z_1) +
4 Z_{o}(2\sqrt{\ap}) +3(4 Z_{o}(2\sqrt{\ap}) - 2Z_2) \right ) \ ,
\nl
Z(\II) &=& \frac{1}{60} \left ( 6(5Z_5-Z_1) + 
10(3Z_3-Z_1)+5(4 Z_{o}(2\sqrt{\ap})-Z_1)+Z_1 \right ) \ .
\label{z-ico}
\ea

There is an interesting relation between the orbifold line at 
radius $R^2=4\a'$ (4-state Potts model or 
$\widehat{SU(2)}_1/{\bf \DD_2}$ model), the tetrahedron and the octahedron, 
that is due to the following chain of normal subgroups:
\be
\mathbb{Z}_2 \times \mathbb{Z}_2 \subset  {\bf T} \subset {\bf O}\ ,
\label{norsub}
\ee
with ${\bf O}/{\bf T}= \mathbb{Z}_2$ and 
${\bf T}/\mathbb{Z}_2 \times \mathbb{Z}_2 =\mathbb{Z}_3 $.
Accordingly, these models are related among themselves by successive 
abelian orbifold operations and backward by simple current extensions. 
Moreover the ${\bf O}$ model can be considered as a 
non-abelian $S_3$ orbifold of the $4$-state Potts model.

The characters of the chiral algebras of the three Ginsparg models
can be found as follows \cite{dvvv}: we express the traces 
$z \left [^g_h \right ]$ in the various orbifold sectors
in terms of $\th$ functions, using the formulae,
\be
Z_n = \frac{1}{n} \sum_{r=0}^{2n-1}\sum_{s=0}^{n-1} 
\left | \th\!\left[^{r/2n}_{s/n} \right]\!(q) \right|^2  \ ,
\label{toi1}
\ee
where the $\th$ functions are defined as follows,
\be
\th\!\left[^{a}_{b} \right]\!\left(q \right)
\equiv
\frac{\Th\!\left[^{a}_{b} \right]\! \left(q^2 \right) }{\h (q)}\ ,
\qquad\qquad
\Th\!\left [^a_b \right ](q) \equiv \sum_{m \in \Z} 
q^{\frac{1}{2}(m-a)^2}e^{-2 \pi i m b} \ .
\label{th-def}
\ee
Then, we can express the orbifold characters using
Eq. (\ref{chi-orb}) and  the character tables 
for $SL_2(\Z_3)$, $GL_2(\Z_3)$  and $SL_2(\Z_5)$ (Appendix \ref{toi-tab}).
The result for the tetrahedron characters  is shown hereafter 
\cite{dvvv}, while the other cases are listed in Appendix \ref{toi-ch}.

$\TT$. The field content of the tetrahedron model consists of 
$21$ chiral fields: $7$ in the untwisted sector, $2$ in the  
$\mathbb{Z}_2$-twisted sector and $6$ for each of the two  
$\mathbb{Z}_3$-twisted sectors.
In the untwisted sector, we find ($ i = 0, 1, 2$):
\be
\begin{array}{llll}
u_i &= \frac{1}{12} \th\!\left[^0_0 \right]
+ \frac{\w^i}{3} \th\!\left[^{\ 0}_{1/3} \right]
+ \frac{\ov{\w}^i}{3} \th\!\left[^{\ 0}_{2/3} \right]
+ \frac{1}{4} \th\!\left[^{\ 0}_{1/2} \right]  \ ,
& &h = 0, 4, 4 \ ,
\\
j &=  \frac{1}{4} \th\!\left[^0_0 \right ] 
- \frac{1}{4} \th\!\left[^{\ 0}_{1/2} \right ]  \ ,
&&h=1 \ ,
\\
\f_i &= \frac{1}{6} \th\!\left[^{1/2}_{\ 0} \right] 
- \frac{\w^{i+2}}{3} \th\!\left[^{1/2}_{1/3} \right]
- \frac{\ov{\w}^{i+2}}{3} \th\!\left[^{1/2}_{2/3} \right]  \ , 
&&h=\frac{1}{4}, \frac{9}{4}, \frac{9}{4}\ ,
\end{array}
\label{tet1}
\ee
where $\w = \exp(2i\pi/3)$.
We can see that the identity representation of  $\widehat{SU(2)}_1$ decomposes
according to the representation of $A_4 \subset SL_2(\Z_3)$ while 
the spin one-half representation according to those  $SL_2(\Z_3)$ 
representations that are projective $A_4$ representations. 
The two characters in the $\mathbb{Z}_2$-twisted sector,
\be
\begin{array}{llll}
\s &= \frac{1}{2} \th\!\left[^{1/4}_{\ 0} \right] + 
\frac{1}{2} \th\!\left[^{1/4}_{1/2} \right] \ ,
&&h=\frac{1}{16} \ ,
\\ 
\t &= \frac{1}{2} \th\!\left[^{1/4}_{\ 0} \right] - 
\frac{1}{2} \th\!\left[^{1/4}_{1/2} \right] \ , 
& &h=\frac{9}{16} \ ,
\end{array}
\label{tet2}
\ee
clearly reflect the structure of their $\mathbb{Z}_2$ stabilizer 
as in the orbifold line. 
Finally the $\mathbb{Z}_3$-twisted characters ($ i = 0, 1, 2$),
\be
\begin{array}{llll}
\w^+_i &=  \frac{1}{3} \th\!\left[^{1/3}_{\ 0} \right] 
+\frac{\w^i}{3} \th\!\left[^{1/3}_{1/3} \right]
+\frac{\ov{\w}^i}{3} \th\!\left[^{1/3}_{2/3} \right] \ , 
&&h=\frac{1}{9}, \frac{4}{9}, \frac{16}{9}\ ,
 \\
\w^-_i &=  \frac{1}{3} \th\!\left[^{2/3}_{\ 0} \right] 
+\frac{\ov{\w}^{i-1}}{3} \th\!\left[^{2/3}_{1/3} \right]
+\frac{\w^{i-1}}{3} \th\!\left[^{2/3}_{2/3} \right] \ , 
&&h=\frac{1}{9}, \frac{4}{9}, \frac{16}{9}\ ,
 \\
\th^+_i &=  \frac{1}{3} \th\!\left[^{1/6}_{\ 0} \right] 
+\frac{\w^i}{3} \th\!\left[^{1/6}_{1/3} \right]
+\frac{\ov{\w}^i}{3} \th\!\left[^{1/6}_{2/3} \right] \ , 
&&h=\frac{1}{36}, \frac{25}{36}, \frac{49}{36}\ ,
 \\
\th^-_i &=  \frac{1}{3} \th\!\left[^{5/6}_{\ 0} \right] 
+\frac{\ov{\w}^{i-1}}{3} \th\!\left[^{5/6}_{1/3} \right]
+\frac{\w^{i-1}}{3} \th\!\left[^{5/6}_{2/3} \right] \ , 
&&h=\frac{1}{36}, \frac{25}{36}, \frac{49}{36}\ ,
\end{array}
\label{tet3}
\ee
organize according to the representations of  $\mathbb{Z}_3$.
From the explicit form of the characters we can calculate the modular 
$S$ matrix. The result (multiplied by $12 \sqrt{2}$) is:
\be
\begin{array}{c|ccccccccc}
 & u_j& j& \f_j& \s& \t& \w_j^+& \w^-_j& \th^+_j&\th^-_j 
\\ \hline 
u_i &1 &3 &2 &6 &6 &4 \w^i & 4 \ov{\w}^i&4 \w^i & 4 \ov{\w}^i
\\
j &3 &9 &6& -6 &-6 &0 &0 &0 &0  
\\
\f_i &2 &6 &-4 &0 &0 &-4 \w^i &- 4 \ov{\w}^i &4 \w^i & 4 \ov{\w}^i
\\
\s &6 &-6 &0 &6\sqrt{2} &-6\sqrt{2} &0 &0 &0 &0 
\\
\t &6 &-6 &0 &-6\sqrt{2} & 6\sqrt{2}&0 &0 &0 &0 
\\
\w^+_i &4 \w^j &0 &-4 \w^j & 0&0 &4 \ov{\a}^2 \ov{\w}^{i+j}&4 \a^2 \w^{i+j} 
&4 \a \w^{2i+j}&4 \ov{\a} \ov{\w}^{2i+j}  
\\
\w^-_i &4 \ov{\w}^j &0 &-4 \ov{\w}^j &0 &0 &4 \a^2 \w^{i+j} 
&4 \ov{\a}^2 \ov{\w}^{i+j} 
&4 \ov{\a} \ov{\w}^{2i+j}   & 4 \a \w^{2i+j}
\\
\th^+_i &4 \w^j &0 &4 \w^j &0 &0 &4 \a \w^{i+2j} &4 \ov{\a} \ov{\w}^{i+2j} 
&4\ov{\b}\w^{i+j} &4 \b \ov{\w}^{i+j} 
\\
\th^-_i &4\ov{\w}^j &0 &4\ov{\w}^j &0 &0 &4 \ov{\a} \ov{\w}^{i+2j} 
&  4 \a \w^{i+2j}
&4 \b \ov{\w}^{i+j}  & 4\ov{\b}\w^{i+j}
\end{array}
\label{tetra-s}
\ee
where $i,j=0,1,2$, $\a = \exp(2 i \pi /9)$ and $\b = \exp(i \pi/9)$.

The complex $S$ matrix implies a non-trivial conjugation for this model. 
Actually there are five self-conjugate fields corresponding to 
$\{ u_0, j, \f_0, \s, \t \}$ while the fields
$\{u_1, \f_1, \w^+_i, \th^+_i \}$ are mapped to 
$\{ u_2, \f_2, \w^-_{i}, \th^-_{i}\}$.
As a consequence there exist both the charge-conjugation and 
the diagonal modular invariant.
The fields $u_i$ form a $\mathbb{Z}_3$ group of simple currents 
that allow the extension of the tetrahedron to the $4$-state Potts model.
Furthermore the untwisted field fusion rules coincide with the 
representation algebra of the group
${\bf T}$ as expected from the general discussion in Ref.\cite{dvvv}.

$\OO$. The octahedron model contains $28$ chiral fields: eight from 
the untwisted sector \\
$\{ u_{\pm}, u_f, j_{\pm}, \f_{\pm}, \f_f \}$, two from the  
$\mathbb{Z}_2$-twisted sector
$\{ \m_r \}$, $r= 0, 1$, six from the  $\mathbb{Z}_3$-twisted sector 
$\{ \w_i, \th_i \}$, $i= 0, 1, 2$,
and twelve from the  $\mathbb{Z}_4$-twisted sector 
$\{ \a_k, \b_k, \s_{\pm}, \t_{\pm} \}$, $k = 0, 1, 2, 3$. 
This notation reflects the $\mathbb{Z}_2$-orbifold relation with 
the ${\bf T}$ model.
Actually the $u_f$ and the $\f_f$ characters originate from 
the identification of the $u_1$, $u_2$ and of the $\f_1$, $\f_2$ 
characters of the tetrahedron whereas 
$u_{\pm}$, $j_{\pm}$, $\f_{\pm}$, $\s_{\pm}$ and $\t_{\pm}$
arise from the splitting of the corresponding 
characters in the ${\bf T}$ model. 
To these five fixed points correspond 10 new 
twisted fields and finally the fields in the two  
$\mathbb{Z}_3$-twisted sectors
are pairwise identified, resulting in only one  $\mathbb{Z}_3$-twisted sector.
The simple current that extends the octahedron model back to 
the tetrahedron is the chiral field $u_-$.
The octahedron $S$ matrix (multiplied by $24\sqrt{2}$) is: 

$$
\begin{array}{c|ccccccccccccc}
 & u_{\pm}& u_f& j_{\pm}& \f_{\pm}&\f_f&\m_s& \s_{\pm}& \t_{\pm}
& \w_j&  \th_j&\a_l&\b_l 
\\ \hline 
 u_{\pm}& 1 &2 &3 &2 &4 &\pm 12  &6  & 6
&8 &8 & \pm 6 & \pm 6 
\\
u_f& 2 & 4 & 6 & 4 & 8 & 0  & 12 & 12
& - 8 & -8 & 0 & 0
\\
j_{\pm}& 3 & 6 & 9 & 6 & 12 & \pm 12  & -6 & -6 
& 0 & 0& \mp 6 & \mp 6 
\\
\f_{\pm}& 2 & 4 & 6 & - 4 & - 8 & 0  & 0 & 0 
& - 8 & 8 & \pm 6 \sqrt{2} & \mp 6 \sqrt{2}
\\
\f_f& 4 & 8 & 12 & -8 & - 16 & 0 & 0 & 0 & 
8 & -8 & 0 & 0 
\\
\m_r& \pm 12 & 0 & \pm 12 & 0 & 0& 12 \sqrt{2}\e_{rs} & 0 & 0
& 0& 0& 0 & 0
\\ 
\s_{\pm}&6 & 12 & -6 & 0 & 0& 0 & 6 \sqrt{2} & - 6 \sqrt{2}
& 0 & 0 & \pm c_l  & \pm s_l 
\\
\t_{\pm}&6 & 12 & -6 & 0 & 0& 0 & -6 \sqrt{2} &  6 \sqrt{2}
& 0 & 0 & \pm s_l  & \mp c_l
\\ 
\w_i& 8 & -8 & 0 & -8 & 8 & 0 & 0 & 0 
& a_{ij} & b_{ij} & 0 & 0
\\
\th_i& 8 & -8 & 0 & 8 & -8 & 0 & 0 & 0 
& b_{ji} & d_{ij} & 0 & 0
\\
\a_k& \pm 6 & 0 & \mp 6 & \pm 6 \sqrt{2} & 0 & 0 & \pm c_k & \pm s_k 
& 0 & 0 &  q_{kl} & r_{kl}
\\
\b_k& \pm 6 & 0 & \mp 6 & \mp 6 \sqrt{2} & 0 & 0 & \pm s_k & \mp c_k 
& 0 & 0 & r_{lk} & s_{kl} 
\end{array}
$$

where we have introduced the matrices,
\ba
\e_{rs}&=& (-1)^{r+s}\ ,\hspace{2cm} 
c_k = (-1)^k 12 \cos(\pi/8)\ ,\hspace{.5cm} 
 s_k = (-1)^k 12 \sin(\pi/8)\ ,
\nl 
a_{ij} &=& 16 \R(\ov{\a}^2\ov{\w}^{i+j}) \ , \hspace{1cm} 
b_{ij}=16 \R(\a\w^{2i+j}) \ , \hspace{1cm} 
d_{ij} = 16 \R(\ov{\b}\w^{i+j}) \ , 
\nonumber \\
q_{kl} &=& 12 \R \left (e^{-\frac{i \pi}{16}} i^{l+k} \right ) \ , 
\hspace{0.5cm}
r_{kl}= 12 \R \left (e^{-\frac{3 i \pi}{16}} i^{l-k} \right ) \ , 
\nonumber \\
s_{kl} &=& 12 \R \left(e^{-\frac{9 i \pi}{16}} (-i)^{l+k} \right ) \ , 
\ea
with indices $i,j=0,1,2$ and $k,l=0,1,2,3$.
In this case the $S$ matrix is real and there is only the 
diagonal modular invariant.

$\II$. Finally the icosahedron field content amounts to $37$ chiral 
fields whose characters are listed in Appendix \ref{toi-ch}
together with the $S$ matrix.
There are nine fields in the untwisted sector $\{ u_i, \f_j \}$, with
$i=0,...,4$ and $j=1,...,4$,
two fields in the $\mathbb{Z}_2$-twisted sector $\{\s,\t \}$, six in the  
$\mathbb{Z}_3$-twisted sector $\{\w_i, \th_i \}$, $i=0,1,2$
and twenty in  the $\mathbb{Z}_5$-twisted sector
$\{ \p_k, \r_k, \l_k, \xi_k \}$, $k=0,...,4$.
In this theory the $S$ matrix is also real and there is  
the diagonal modular invariant only.
Furthermore, there are no simple currents among the chiral fields 
of the icosahedron model, in agreement with the fact that 
it can not be obtained through a sequence of abelian orbifold
operations as the ${\bf T}$ and the ${\bf O}$ models.

In studying these orbifolds one should also 
allow for the presence of discrete torsion \cite{vafa}; 
the different possibilities are 
classified by $H^2(G,U(1))$ that, in our case, is non-trivial and 
equal to $\Z_2$ for $G \ = \ \DD_{2n}, \TT , \OO ,\II $. 
Nevertheless, the $\DD_{2n}$ orbifolds
with or without discrete torsion are equivalent \cite{dgh},
and the same is expected for the $\TT , \OO ,\II $ models.


\subsection{Boundary states}

In the previous Section we have discussed 
the chiral sectors of the ${\bf T}, \ {\bf O}$ and ${\bf I}$ orbifold models. 
The boundary coefficients for the charge-conjugation modular invariant can
be read from the respective $S$ matrices according to the Cardy formula
(\ref{cardy-b}). 
The result shows some interesting features:
let us consider the tetrahedron, for example, Eq. (\ref{tetra-s}).
Firstly, we can observe that boundary states corresponding 
to a given twisted field are uncharged with respect to fields in 
different twisted sectors: actually, the $S$ matrix vanishes on
the corresponding block entries.

Secondly, the untwisted boundary states can again be interpreted 
as fractional branes. 
The first four states of the tetrahedron, $|u_i\ra$, $|j\ra$, $i=0,1,2$,
come from the splitting of the D0 brane at the north pole of
$\widehat{SU(2)}_1$ while the next three, $|\f_i\ra$, come from the D0 brane
at the south pole. 
The fractional nature of these branes is confirmed by the number of
possible marginal deformations, respectively 0, 2 and 1, thus showing
that the number of directions of motion is lower than the 3 displacements
of the original D0 brane in the $SU(2)$ three-sphere.
The boundary operator content is given by the annulus amplitudes ($i=0,1,2$):
\ba
A_{u_i,u_i^*} &=&  u_0 \ , \nl 
A_{j,j} &=&  u_0 +u_1 +u_2 + 2\ j \ , \nl
A_{\f_i,\f_i^*} &=&  u_0 + j \ , \nl
A_{\s,\s} = A_{\t,\t} &=& \sum_{i=0}^2 u_i + j 
+ \sum_{i=0}^2 \f_i + 2 \s + 2 \t  \ , \nl
A_{\w^\pm_i,\w^\mp_i} = A_{\th^\pm_i,\th^\mp_i} &=& u_0 + j +\s + \t \ . 
\label{t-ann}
\ea
These amplitudes and the boundary states associated to the twisted fields 
$|\s\ra$, $|\t\ra$, $|\w^\pm_i\ra$ and $|\th^\pm_i \ra$, $i=0,1,2$,
could be further understood by studying 
the geometry of the orbifold space. 
The Cardy states of the octahedron and icosahedron models
present a similar pattern of fractional and twisted branes.

Let us now discuss 
the boundary states for the diagonal modular invariant of the tetrahedron:
there are five Ishibashi states corresponding to the self-conjugate
fields $\{u_0,j,\f_0,\s,\t \}$,
and therefore we expect five boundary states. 
We use the same method applied to the diagonal modular 
invariant of the compactified boson, namely we obtain them from the octahedron 
by a simple current extension, the relevant integer-spin current being
$J=u_-$.
Actually, the $\Z_2$ symmetry ${\cal R}$ in ${\bf O} = {\bf T}/{\cal R}$ 
induces as automorphism of the tetrahedron fusion rules precisely 
the charge conjugation. 
The boundaries for the tetrahedron with conjugation modular invariant 
that preserve only the $\TT/{\cal R}$ orbifold subalgebra then coincide with 
the symmetry preserving boundaries for the theory with diagonal 
modular invariant.

Under the action of $J$, the ten octahedron boundary states that 
correspond to the fields in the ${\cal R}$-twisted sector combine in pairs. 
Explicitly $J$ maps the boundaries $|\m_0 \ra$,  $|\a_0 \ra$, 
$|\a_1 \ra$,  $|\b_0 \ra$ and  $|\b_1 \ra$ respectively to 
$|\m_1 \ra$,  $|\a_2 \ra$, $|\a_3 \ra$,  $|\b_2 \ra$ and $|\b_3 \ra$. 
The boundary coefficients are: 
\be
B_{a i} = \frac{S_{a, i} + S_{J(a), i}}{\sqrt{2 S_{0, i}}} \ ,
\ee
where $S_{i j}$ is the $S$ matrix of the octahedron. 
The resulting boundary coefficients are non zero  
only for the following five combinations of octahedron Ishibashi states:
\ba
& & |u_+ \rra - |u_- \rra \ , \hspace{1cm} 
|j_+ \rra - |j_- \rra \ ,  \hspace{1cm} 
|\f_+ \rra - |\f_- \rra \ , 
\nonumber \\
& & |\s_+ \rra - |\s_- \rra \ ,  \hspace{1cm} 
|\t_+ \rra - |\t_- \rra \ , 
\label{td-bs}
\ea
that are precisely the ${\bf T}$ Ishibashi states resulting from the 
gluing $\W {\cal R}$. 
In this basis the reflection coefficients are:
\be
R = \frac{1}{2\sqrt{2}} \left(
\begin{array}{ccccc}
1 & -1 & \sqrt{2} & \sqrt{2+\sqrt{2}} &\sqrt{2-\sqrt{2}} \\
1 & -1 &  \sqrt{2} & - \sqrt{2+\sqrt{2}} &- \sqrt{2-\sqrt{2}} \\
1 & -1 & - \sqrt{2} & \sqrt{2-\sqrt{2}} &- \sqrt{2+\sqrt{2}} \\
1 & -1 & - \sqrt{2} & - \sqrt{2-\sqrt{2}} & \sqrt{2+\sqrt{2}} \\
2 & 2 & 0 & 0& 0 
\end{array} \right)\ .
\label{td-s}
\ee
As already said,
we can interpret these states as the five symmetry-preserving boundaries 
for the tetrahedron diagonal invariant.
We have explicitly verified that the annulus amplitudes for the 
five boundaries in (\ref{td-s}) are consistent and that the coefficient 
$A^i_{a b}$ give a five dimensional representation of the fusion algebra
(reported in Appendix \ref{toi-ch}).
We have also verified that these boundaries, 
interpreted as symmetry breaking boundaries
for the tetrahedron with conjugation modular invariant, 
have consistent overlaps with the usual Cardy
states and that the corresponding annulus amplitudes contain, as expected, 
the twisted characters of the octahedron. 
For instance two of the ${A}_{i,a}$, $a = 1, ..., 5$ are:
\be
{A}_{u_0,1} = \a_0 + \a_2 \ , \qquad\qquad
{A}_{j,5} = 3 \m_0 + 3\m_1 \ . 
\ee

Finally, the Klein and M\"obius amplitudes for the $\TT-\OO-\II$ models,
in particular those for the diagonal $\TT$ model, are discussed
in Appendix \ref{klein}.


\section{$c=3/2$ superconformal field theories}


\subsection{Moduli space}

The partition functions of $N=1$ superconformal theories
can be written in general as follows \cite{supercft}:
\be
Z = \frac{1}{2} \left   (Z_{NS} + Z_{\widetilde{NS}} + 
Z_{R} \pm Z_{\widetilde{R}} \right ) \ ,
\ee 
where the four terms 
correspond to antiperiodic ($a$) and periodic ($p$) boundary conditions 
for the supercurrent $G$ along the two non-trivial cycles of the torus: 
respectively, $(a,a)$, $(p,a)$, $(a,p)$ and $(p,p)$. 
The last term is the Witten index $Z_{\widetilde{R}}=Tr_R(-1)^F$;
the two choices of the sign are related by
the $\mathbb{Z}_2$ symmetry $(-1)^{F_s}$, that takes the value $+1$ on states
in the NS-NS sector and the value $-1$ on states in the R-R sector. 
Actually, one theory is the orbifold of the other by $(-1)^{F_s}$. 

The simplest realization of superconformal symmetry at $c=3/2$ is 
given by the theory of a free $N=1$ superfield, made by the 
boson field $X$ compactified on a circle of radius $R$
and by the Majorana fermion $\psi$. 
The partition function for this system is the product of 
the familiar lattice sum (\ref{z-circ}) for the boson, and of 
the fermion partition function summed over the spin structures:
\be
Z_{c}(R) =  \sum_{n,m \in \Z} \G_{n,m}  
\left( |o|^2 + |v|^2 + |s|^2  \right ) \ .
\ee
The fermion contribution is expressed in terms of the characters 
of the Ising model \cite{rev-cft}:
\be
\begin{array}{lll}
o = \frac{1}{2} 
\left ( \sqrt{\frac{\th_3}{\h}} + \sqrt{\frac{\th_4}{\h}} \right ) \ , 
&& h=0\ ,
\\
v = \frac{1}{2} 
\left ( \sqrt{\frac{\th_3}{\h}} -  \sqrt{\frac{\th_4}{\h}} \right ) \ , 
&& h=\frac{1}{2}\ ,
\\
s =  \sqrt{\frac{\th_2}{2 \h}}  \ ,
&& h=\frac{1}{16}\ .
\end{array}
\label{is-char}
\ee
whose $S$ matrix is
\be
S = \frac{1}{2}  \left( 
\begin{array}{ccc}
1 & 1 & \sqrt{2}  \cr 
1 & 1 & - \sqrt{2} 
\cr \sqrt{2} & - \sqrt{2} & 0 
\end{array} \right) \ .
\label{is-s}
\ee 

The free superfield compactified on a circle describes
the first family of superconformal field 
theories, parametrized by the radius $R$. These theories possess 
one R-R ground state and the Witten index vanishes. 
Again, the radii $R$ and $\ap/R$ are related by $T$-duality.
At the self-dual radius $R = \sqrt{\ap}$ there is an $N=3$ 
superconformal algebra, resulting 
from the combination of the affine $\su2_1$ symmetry with 
the $N=1$ superconformal symmetry.

The moduli space of $c=3/2$ superconformal theories 
has been investigated in Ref.\cite{dgh}
by identifying the discrete symmetries of a given family of models 
and by building new models by various orbifold constructions.
Let us summarize these results. The discrete symmetries present at
generic values of the circle radius are:
the reflection,
\be
{\cal P}: \hspace{1cm} X \mapsto - X \ , \hspace{1cm} \psi \mapsto - \psi \ ,
\ee  
and the previously mentioned $(-1)^{F_s}$. 
Furthermore, one can identify $X$ modulo translations 
by integer fractions of the compactification radius,
\be
\d_n: \hspace{1cm} X \sim X + \frac{2 \p R}{n} \ , 
\hspace{1cm} \psi \mapsto  \psi \ .
\ee
The resulting model is again the compactified superfield theory 
at radius $R/n$. 
New models were obtained by combining $\d_2$ and the other two involutions. 
As shown in \cite{dgh}, the relevant cases are 
${\cal P}$, $(-1)^{F_s}{\cal P}$ and  $(-1)^{F_s} \d_2$.

The $\Z_2$ orbifold by the symmetry ${\cal P}$ is described by the superfield 
compactified on the interval of length $\p R$, with
partition function:
\be
Z_{o}(R) = \frac{1}{2} 
\left ( \sum_{n,m \in \Z} \G_{n,m} + \left|\sqrt{\frac{2 \h}{\th_2}}\right|^2 
+ \left|\sqrt{\frac{2 \h}{\th_4}}\right|^2 + 
\left|\sqrt{\frac{2 \h}{\th_3}}\right|^2 \right ) 
\left (  |o|^2 + |v|^2 + |s|^2  \right ) \ .
\ee
The T-duality is $Z_{o}(R)=Z_{o}\left (\ap/R\right )$ ; 
the circle line meets the orbifold line at the self-dual radius, 
$Z_{o}(\sqrt{\ap})=Z_{c}(\sqrt{4\ap})$. 
Theories on the orbifold line possess  three R-R ground states,
$u_+s \ov{u_+s}$, $\s_0 o \ov{\s_0o}$ and $\s_1 o \ov{\s_1 o}$ 
and Witten index equal to three.

Using $(-1)^{F_s}{\cal P}$ instead of ${\cal P}$, one obtains a very similar 
model that is nothing else than the orbifold by
$(-1)^{F_s}$ of the previous theory, called 
the orbifold-prime theory \cite{dgh}. Actually, the simple current
$u_{-}v$ of the orbifold theory implements the $(-1)^{F_s}$ symmetry;
the partition function reads: 
\ba
Z_{o'}(R) &=&  \frac{1}{2}  \left ( \sum_{n,m \in \Z} \G_{n,m} 
+ \left|\sqrt{\frac{2 \h}{\th_2}}\right|^2 \right )
\left (  |o|^2 + |v|^2  \right ) 
+ \frac{1}{2}  \left ( \sum_{n,m \in \Z} \G_{n,m} - 
\left|\sqrt{\frac{2 \h}{\th_2}}\right|^2 \right )|s|^2 
\nonumber \\ 
&+& \frac{1}{2} \left ( \left|\sqrt{\frac{2 \h}{\th_4}}\right|^2 
- \left|\sqrt{\frac{2 \h}{\th_3}}\right|^2 \right ) 
\left ( o \ov{v} + v \ov{o}  \right ) 
+ \frac{1}{2} \left ( \left|\sqrt{\frac{2 \h}{\th_4}}\right|^2 
+ \left|\sqrt{\frac{2 \h}{\th_3}}\right|^2 \right )|s|^2 \ .
\ea
At the rational points $R=\sqrt{\ap k}$, this 
partition function can be rewritten in terms of the characters
of the rational $c=1$ $S^1/\Z_2$ orbifold (\ref{z2-chi}):
\ba
Z_{o'}(\sqrt{\a'k}) &=&  \left (|u_+|^2+|u_-|^2 + |\f_+|^2 +|\f_-|^2\right ) 
\left (  |o|^2 + |v|^2  \right ) 
\nonumber \\ 
&+& \left (u_+\ov{u}_- + u_-\ov{u}_+ + \f_+\ov{\f}_- + \f_-\ov{\f}_+ 
  \right ) |s|^2 
+ \sum_{r=1}^{k-1} |\chi_r|^2 \left (  |o|^2 + |v|^2 + |s|^2  \right )  
\nonumber \\ 
&+& \sum_{i=0}^1 \left [ \left ( \s_i \ov{\t}_i + \t_i \ov{\s}_i \right ) 
\left ( o \ov{v} + v \ov{o}  \right )  
+ \left ( |\s_i|^2+|\t_i|^2  \right ) |s|^2 \right ] \ .
\label{z-op}
\ea
From this expression, it is clear that there are no R-R ground states 
since the characters
$u_+s$, $\s_0 o$ and $\s_1o$ now appear in off-diagonal combinations.

The orbifold of the circle theory by $(-1)^{F_s} \d_2$ 
yields the so-called super-affine line, whose partition function is:
\ba 
Z_{sa}(R) &=&  \sum_{n,m \in \Z} \G_{2n,m} \left (  |o|^2 + |v|^2  \right ) 
+ \sum_{n,m \in \Z} \G_{2n+1,m}  |s|^2  
\nl
&+& \sum_{n,m \in \Z} \G_{2n+1,m+\frac{1}{2}}
\left ( o \ov{v} + v \ov{o}  \right ) 
+\sum_{n,m  \in \Z} \G_{2n,m+\frac{1}{2}} |s|^2 \ .
\ea
The rational theories at radii $R=\sqrt{2 \ap m}$ display $12m$ sectors and  
their partition functions can be written, for $m$ even
($r=-2m+1,...,2m$):
\ba
Z_{sa}\left(\sqrt{2 \ap m}\right) &=&  
\sum_{r \ even } |\chi_r|^2  \left (  |o|^2 + |v|^2  \right ) 
+ \sum_{r \ odd } |\chi_r|^2  |s|^2 \nonumber \\
&+& \sum_{r \ even } \chi_r \ov{\chi}_{r+2m} |s|^2  
+  \sum_{r \ odd }  \chi_r \ov{\chi}_{r+2m}
\left ( o \ov{v} + v \ov{o}  \right ) \ ,
\label{sa-even}
\ea
and for $m$ odd,
\ba
Z_{sa}\left(\sqrt{2 \ap m}\right) &=& 
\sum_{r \ even} |\chi_r o + \chi_{r+2m}v|^2 + 
\sum_{r=1, \ odd}^{2m-1} |( \chi_r  + \chi_{r+2m})s|^2   \ .
\label{sa-odd}
\ea
These modular invariants are easily understood 
noticing that they can be obtained as simple current constructions of the
circle theory.
The simple current is the field $\chi_{2m}v$ of the circle 
theory whose conformal dimension is $h = (m+1)/2$ and hence we have an 
automorphism or extension modular invariant for
 $m$ even or odd, respectively.
Note that the rational partition functions (\ref{sa-even}), (\ref{sa-odd})
are written for the diagonal pairing of charges;
one should also bear in mind the analogous expressions with 
charge-conjugation pairing, e.g. $|\chi_r|^2 \to \chi_r \ov{\chi}_{-r}$.

The T-duality of the super-affine line is 
$Z_{sa}(R)=Z_{sa}\left (2 \ap/R \right)$. 
At the self-dual point $R=\sqrt{2 \ap}$ there is a super-affine 
$\widehat{SO(3)}_1$ symmetry and the partition function 
can be rewritten in terms of the affine characters:
\ba
O_3 &=& \frac{1}{2} \left [ \left (\frac{\th_3}{\h} \right )^{3/2} 
+  \left ( \frac{\th_4}{\h} \right )^{3/2} \right ] \ , 
\nl
V_3 &=& \frac{1}{2} \left [ \left (\frac{\th_3}{\h} \right )^{3/2}  
-   \left ( \frac{\th_4}{\h}  \right )^{3/2} \right ] \ , 
\nl
S_3 &=&  \frac{1}{\sqrt{2}}
\left ( \frac{\th_2}{ \h}  \right )^{3/2} \ .
\ea
There are no R-R ground states on the super-affine line. 
The super-affine and the orbifold-prime lines intersect at the point:
$Z_{sa}(\sqrt{4\ap})=Z_{o'}(\sqrt{\ap})$.

The fifth and final family of theories is called 
the super-orbifold line and is obtained  by
orbifolding the super-affine line by ${\cal P}$ \cite{dgh}.
Let us present the partition function directly at the rational points 
$R=\sqrt{2 \ap m}$:
they are simple current modular invariants of the orbifold, 
the simple current being $\f_+ v$ with conformal
weight $h = (m+1)/2$. As for the super-affine line we have 
to distinguish between even and odd values of $m$: 
the first case gives the automorphism modular invariants, 
\ba
Z_{so}\left(\sqrt{2\ap m}\right) &=& 
 \left (|u_+|^2+|u_-|^2 + |\f_+|^2 +|\f_-|^2\right ) 
\left (  |o|^2 + |v|^2  \right ) 
\nonumber \\ 
&+&  
\left (u_+\ov{\f}_+ + \f_+\ov{u}_+ + u_{-}\ov{\f}_{-} + 
\f_{-}\ov{u}_{-}   \right ) |s|^2 
+ \sum_{r=1}^{m-1} |\chi_{2r}|^2 \left (  |o|^2 + |v|^2  \right )  
\nonumber \\ 
&+&  \sum_{r=1}^{m} |\chi_{2r-1}|^2|s|^2 
+ \sum_{r=1}^{m-1} \chi_{2r} \ov{\chi}_{2m-2r}|s|^2  
+ \sum_{r=1}^{m} \chi_{2r-1} \ov{\chi}_{2m-2r+1}
\left ( o \ov{v} + v \ov{o}  \right ) 
\nonumber \\
&+&  \left ( |\s_0|^2+|\t_0|^2  \right ) 
\left (  |o|^2 + |v|^2 +  |s|^2  \right ) 
\nonumber \\
&+&  \left ( \s_1 \ov{\t}_1 + \t_1 \ov{\s}_1 \right ) 
\left ( o \ov{v} + v \ov{o}  \right )  
+ \left ( |\s_1|^2+|\t_1|^2  \right ) |s|^2  \ .
\label{z-so1}
\ea
The odd $m$ case gives extension modular invariants:
\ba
Z_{so}\left(\sqrt{2\ap m}\right)&=&   
|u_+o+\f_+v|^2+|\f_+o+u_+v|^2+|u_-o+\f_-v|^2+|\f_-o+u_-v|^2 
\nonumber \\
&+&  \sum_{r=1}^{m-1} | \chi_{2r}o + \chi_{2m-2r}v|^2
+ \sum_{r=1}^{\frac{m-1}{2}} |\chi_{2r-1} + 
\chi_{2m-2r+1}|^2|s|^2+2|\chi_m s|^2 
\nonumber \\
&+& |\s_1o+\t_1v|^2+|\t_1o+\s_1v|^2+2(|\s_0|^2+|\t_0|^2)|s|^2 \ .
\label{z-so2}
\ea
Theories along this line possess one R-R ground state.
The super-orbifold line crosses both the orbifold-prime and the circle line: 
$Z_{so}(\sqrt{4\ap})=Z_{o'}(\sqrt{4\ap})$ 
and $Z_{so}(\sqrt{2\ap})=Z_{c}(\sqrt{2\ap})$.

\begin{figure}[ht]
\begin{center}
\input{moduli.pstex_t}
\end{center}
\caption{The continuous lines of $c=3/2$ superconformal theories:
the values of the compactification radii $R_c,R_o,R_{o'},R_{so}$ 
and $R_{sa}$, are shown for $\ap=1/2$: they parametrize
the circle, orbifold, orbifold-prime, super-orbifold 
and super-affine theories, respectively \cite{dgh}.}
\label{mod-fig}
\end{figure}
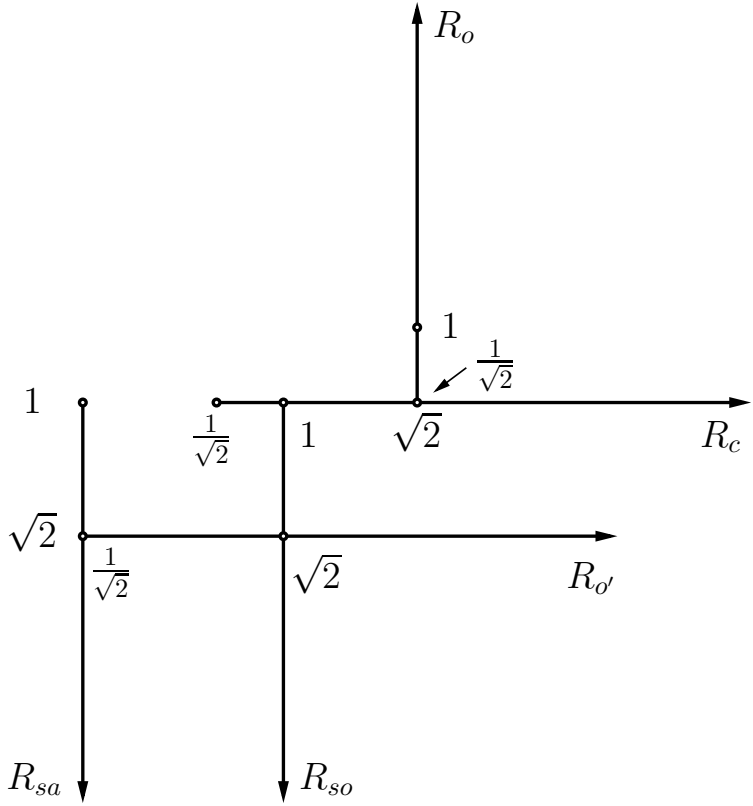

The five lines of theories are schematically drawn in Fig.\ref{mod-fig}
\cite{dgh}. The pattern is more easily understood
in terms of the chiral algebras underlying the various families
of rational theories, at the particular rational points considered before.
There are only two independent chiral algebras:
the direct product of the Ising-model algebra with either the
$\widehat{U(1)}_k$  or the $\widehat{U(1)}_k/\mathbb{Z}_2$ algebra.
The first one describes the circle and super-affine lines, 
the second applies to the orbifold, the orbifold-prime and 
the super-orbifold lines.
Different lines for the same chiral algebra correspond to 
modular invariants generated by $\Z_2$ simple currents: 
the current $\c_{2m} v$ of the circle theories yields 
the super-affine line, and the currents $u_- v$ and $\f_+ v$
of the orbifold algebra produce the orbifold-prime and 
the super-orbifold lines, respectively.

We now consider orbifolds by discrete symmetries that exist 
at particular points on the moduli space \cite{dgh}. 
One natural possibility is represented by the self-dual point along the
circle line: modding out $Z_c(\sqrt{\ap})$  
by the $SU(2)$ discrete subgroups, one finds 
the product of the corresponding $c=1$ theories times the Majorana fermion. 
More interesting $N=1$ superconformal models can be constructed 
as orbifolds of the super-affine theory
at the self-dual point $R=\sqrt{2 \ap}$, that displays 
the $\widehat{SO(3)}_1$ symmetry. 
Actually, this theory can be realized by three Majorana fermions
$\psi^i$, $i=1,2,3$, and the supercurrent
$G = - \frac{1}{12} \e_{ijk} \psi^i \psi^j \psi^k$ is $SO(3)$ invariant;
thus, the quotient by  the discrete subgroups of $SO(3)$ does not spoil
the superconformal symmetry.
The $\CC_n$-orbifolds yield  points along either the super-affine or 
the circle line (we set $2 \ap = 1$ here):  
\be 
Z(\CC_{2n+1}) = Z_{sa}\left({2n+1} \right) \ , \hspace{1cm} 
Z(\CC_{2n}) = Z_{c}\left({n/2}\right) \ .
\label{z-scyc}
\ee
One similarly finds that:
\be 
Z(\DD_{2n+1}) = Z_{so}\left({2n+1} \right) \ ,\quad 
Z(\DD_{2n})   = Z_{o}\left({n/2}\right) \ , \quad
Z(\DD'_{2n})  = Z_{o'}\left({n/2}\right) \ .
\label{z-sdied}
\ee
The second theory for the $\DD_{2n}$ orbifolds, namely $Z( \DD'_{2n})$,
is found by introducing a $\mathbb{Z}_2$ discrete torsion 
(in the following, the models with discrete torsion will be 
labeled by a prime).
Finally, the orbifolds of the self-dual super-affine theory 
by the non-abelian groups $A_4$, $S_4$ and $ A_5$
produce six isolated points: $\TT,\OO,\II$ and $\TT',\OO',\II'$ that 
will be discussed in Section 6.


\subsection{Boundary states for the superconformal lines}

The boundary states for the circle line are easily obtained 
as combinations of the boundary states 
for the compactified boson and the Majorana fermion 
(see Eqs.(\ref{DN-bound}),(\ref{is-s})). 
Restricting ourselves to the rational case $R^2 = \ap k$, we have: 
\ba
|r,o \ra_c &=& \frac{1}{(8k)^{1/4}} 
\sum_{\ell=-k+1}^{k} e^{-\frac{i \pi rl}{k}} 
\left ( |\ell\rra |o \rra + |\ell\rra |v \rra + 
\sqrt{2} |\ell\rra |s \rra \right ) \ , 
\nonumber \\
|r,v \ra_c &=& \frac{1}{(8k)^{1/4}} \sum_{\ell=-k+1}^{k} 
e^{-\frac{i \pi rl}{k}} 
\left ( |\ell\rra |o \rra + |\ell\rra |v \rra - 
\sqrt{2} |\ell\rra |s \rra \right ) \ , 
\nonumber \\
|r,s \ra_c &=& \frac{1}{(2k)^{1/4}} 
\sum_{\ell=-k+1}^{k} e^{-\frac{i \pi rl}{k}} 
\left ( |\ell\rra |o \rra - |\ell\rra |v \rra  \right ) \ ,
\label{bdc1}
\ea
where $ |\ell\rra |o \rra$,  $ |\ell\rra |v \rra$ and $ |\ell\rra |s \rra$, 
are the products of Ishibashi states for the rational boson and 
the Ising model (there are $6k$ boundary states in total).
According to the values of the boundary coefficients for the R-R fields,
the three types of boundary states in (\ref{bdc1}) can be considered
as positively charged, negatively charged and uncharged 
boundary states, respectively.

On the super-affine line at $R^2 = \ap k$ with $k=2m$, 
one expects $3m$ boundary states: they can be obtained acting on the 
boundaries of the circle with the operation $(-1)^{F_s} \d_2$ or 
equivalently with the simple current $J= \chi_{2m} v$,
that is freely acting.
Therefore, there are neither fixed points nor fractional branes and 
the invariant boundary states are made of pairs, as follows:
\ba
|r,o \rangle_{sa} &=& \frac{1}{\sqrt{2}} 
\left ( |r,o \ra_c +|r+k, v \ra_c  \right ) \ , \hspace{1cm} 
r = 0, ..., 2k-1,  
\nonumber \\
|r,s \rangle_{sa} &=& \frac{1}{\sqrt{2}} 
\left ( |r,s \ra_c +|r+k, s \ra_c  \right ) \ , \hspace{1cm} 
r = 0, ..., k-1  \ .
\label{bdc2}
\ea
For $m$ odd the bulk modular invariant is of extension type and 
 the boundaries $|r,o \ra_{sa} $ with $r$ 
odd are breaking the extended symmetry.

The Cardy boundary states of the orbifold line are again products of 
boundary states for the bosonic orbifold with states for the Ising model, 
for a total of  $3(k+7)$ boundary states.
They are labeled by the corresponding fields:
in the untwisted NS sector, there are $u_\pm I, \f_\pm I, \c_r I$,
with $I=o,v$; in the untwisted R sector, $u_\pm s, \f_\pm s, \c_r s$;
in the twisted NS sector, $\s_i s, \t_i s$, with $i=0,1$;
finally, in the twisted R sector, $\s_i I, \t_i I$, $I=o,v$.
We can distinguish them according to their R-R charges,
taking into account that in this case both untwisted and twisted
charges appear.
For instance, the states $|u_\pm I\ra$, have 
both untwisted and twisted charges, while the states 
$|u_\pm s\ra$ carry only twisted charges.

Starting from this set of boundary states and acting with 
the $\Z_2$ simple current $J = \f_+ v$ we can obtain the boundary states 
for the super-orbifold line.
From the partition function (\ref{z-so1},\ref{z-so2}), we expect 
$(3m + 15)$ Ishibashi states at radius $R^2 = \ap k= 2 \ap m $. 
Under the action of the simple current, $6(m + 3)$ boundaries 
are combined in pairs, while 3 of them are fixed, leading to
the required $3m+15$ boundary states. 
In order to construct these boundary states, we must know 
the orbits of the simple current: the three fixed points
are $\chi_m s$, $\s_0 s$ and $\t_0 s$; the representations,
\be
u_{\pm}o, \ \ \f_{\pm}o, \ \ \chi_{r}o, \ \ \s_0o, \ \ 
\t_0o, \ \ \s_1o, \ \ \t_1o  \ , \qquad\qquad r=1,\dots,2m-1\ ,
\ee
are respectively paired with the representations,
\be
\f_{\pm}v, \ \ u_{\pm}v, \ \ \chi_{2m-r}v, \ \ \s_0v, \  
\ \t_0v, \ \ \t_1v, \ \ \s_1v \ , 
\ee
while the representations, 
\be
u_{\pm}s, \ \ \chi_{r}s, \ \ \s_1s \ ,\qquad\qquad r=1, ..., m-1 \ ,
\ee
are paired with
\be
\f_{\pm}s, \ \ \chi_{2m-r}s, \ \ \t_1s \ .
\ee

Again the boundary states of the super-orbifold follow, for $m$ even, 
the general pattern
described in \cite{ca} for automorphism modular invariants 
generated by a $\mathbb{Z}_2$ current of 
half-integer spin. In particular we have a set of $3m+9$ 
invariant boundaries in one-to-one
correspondence with the length-two orbits of the simple current,
as in Equation (\ref{inv-b}).
In addition there are six fractional boundary states, 
two for each of the fixed points of the simple current.
These boundary states have the form displayed in (\ref{frac-b})
with the fixed-point $\widetilde{S}$-matrix equal to that of the
Ising model (\ref{is-s}) in the basis 
$\{|\s_0 \rra |s \rra, |\t_0 \rra |s \rra, |\chi_m \rra |s \rra \}$.

Finally the boundary states for the orbifold-prime line can be obtained 
acting with the simple current $J=u_- v$ on the orbifold states. 
All the states labeled by NS fields are paired 
and give $(k+5)$ boundary states, 
while among those labeled by Ramond fields, 
12 are paired and $(k-1)$ are fixed. 
The total number of boundaries is $(3k+9)$, in agreement with the
number of Ishibashi states (see Eq.(\ref{z-op})).

The $(k-1)$ fixed boundary states,
\ba
|r,s \rangle_{o} &=& \left ( \frac{2}{k} \right )^{1/4}  
\Bigg [ |u_+ \rra + |u_- \rra 
+ (-1)^r|\f_+ \rra + (-1)^r|\f_- \rra  
\nonumber \\
&+&  \left. \sum_{l=1}^{k-1}\sqrt{2} 
\cos{ \left (\frac{\pi r l}{k} \right )}|l \rra  \right] 
\left ( |o \rra - |v \rra  \right ) \ , 
\hspace{1cm} r=1, ..., k-1,
\ea
give rise to new boundaries that differ by their charges with respect
to the $(k-1)$ R-R fields $\chi_r s$, with $r=1, ..., k-1$:
\ba
|r, s, \ \pm \rangle_{o'} 
&=& \frac{1}{(2k)^{1/4}}  \Bigg [
 |u_+ \rra + |u_- \rra 
+ (-1)^r|\f_+ \rra + (-1)^r|\f_- \rra 
\nonumber \\
&+& \left. \sum_{l=1}^{k-1}\sqrt{2} 
\cos{\left ( \frac{\pi r l}{k} \right )}|l \rra  \right] 
\left ( |o \rra - |v \rra  \right ) 
\nonumber \\
&\pm& 
\frac{1}{(2k)^{1/4}} \sum_{l=1}^{k-1} 2 
\sin{\left (\frac{\pi r l}{k} \right )}|l \rra 
|s \rra  \ ,  \hspace{1cm}  r=1, ..., k-1 \ .
\ea
One can check that these states give consistent annulus amplitudes. 


\section{Superconformal $\TT$-$\OO$-$\II$ models}

In this Section we find the chiral fields
for the superconformal  $\TT$-$\OO$ models  
and then describe their boundary states.
The discussion parallels that of the three $c=1$ models:
we start from the self-dual point on the super-affine
line and mod it by the symmetry groups $A_4$, $S_4$ and $A_5$.
Taking into account the discrete torsion, one obtains three further models,
$\TT'$, $\OO'$, $\II'$, that can also be realized as  $(-1)^{F_s}$ 
orbifolds of the torsionless $\TT$, $\OO$, $\II$ models
(recall that $(-1)^{F_s}$ also relates
the orbifold and the orbifold-prime line).
Note that the two triples differ in the number of R-R ground states.

The  $\TT$ model can also be realized as a  $\mathbb{Z}_3$ orbifold
of the theory made by the product of three Ising models,
that is found  at $R=\sqrt{2\a'}$ on the orbifold line.
From the chain of inclusions (\ref{norsub}),
it is then clear that the $\OO$ model can be obtained as a permutation
orbifold of the triple Ising theory, ${\it i.e.}$ as a non-abelian
${\bf S}_3$ orbifold:
${\rm (Ising)}^{\otimes 3}/\Z_3 = \TT$ and 
${\rm (Ising)}^{\otimes 3}/S_3 = \OO$.
General expressions for permutation orbifolds
have been given in Ref.\cite{bantay}, and agree with our findings.
 
The characters and $S$ matrices for the superconformal $\TT$ and $\OO$
models are reported in Appendix \ref{stoi-ch}; they are obtained as in
the bosonic case (\ref{z-tet}-\ref{z-ico}), 
by first expanding the partition functions in 
terms pertaining to the mutually commuting subgroups of $\TT-\OO$
(\ref{z-scyc},\ref{z-sdied}), and then by expressing the latter
in terms of $\Th$ functions (Eq.(\ref{th-def})), 
using the identities (for $R^2=2 \ap n$):
\ba
\!\!\!\!\sum_{p,w\in \Z} \frac{1\pm(-1)^p}{2}\G_{p,w} 
&=& \frac{1}{2n|\eta(q)|^2} \sum_{r=0}^{2n-1}\sum_{s=0}^{2n-1} 
\frac{1\pm(-1)^r}{2}
\left |\Th\!\left[^{r/2n}_{s/2n} \right](q) \right|^2  \ ,
\label{s-theta} \\
\!\!\!\!\sum_{p,w \in \Z} \frac{1\pm(-1)^p}{2}\G_{p,w+\frac{1}{2}} 
&=& \frac{1}{2n|\eta(q)|^2} \sum_{r=0}^{2n-1}\sum_{s=0}^{2n-1} 
\frac{1\pm(-1)^{r+n}}{2}(-1)^s
\left | \Th\!\left[^{r/2n}_{s/2n} \right](q) \right|^2 .
\nonumber
\ea

Let us now describe the modular invariant partition functions and the
corresponding boundaries for these models.

{\bf T}. The field content of the superconformal $\TT$ model
consists of 35 fields: 22 of them belong to the
NS sector and 13 to the R sector (see Appendix \ref{stoi-ch}
for notations and explicit expressions). 
There are several modular invariants: first of all the $S$ matrix 
is complex and therefore the conjugation and diagonal invariants are distinct. 
Moreover, starting from the charge-conjugation modular invariant 
$Z({\bf T}_c)$, we can obtain three further modular invariants using
two $\mathbb{Z}_2$ operations.
The first is  the orbifold 
by $(-1)^{F_s}$: this is obtained through the action of the simple current $\x_0$,
that is the primary field containing the supercurrent, as usual. 
The second is the exchange $\s_v \leftrightarrow \t_o$, that is 
an automorphism of the fusion rules.
The resulting partition functions are:
\ba
Z({\bf T}_c) &=& 
\left ( |\s_o|^2 + |\s_v|^2 + |\t_o|^2 + |\t_v|^2 \right ) 
+ \sum_i \chi_i \ov{\chi}_{i^*} \ , 
\nl
Z({\bf T}_{ca}) &=& 
\left ( |\s_o|^2 + \s_v \overline{\t_o}+ \t_o \overline{\s_v} + 
|\t_v|^2 \right ) + \sum_i \chi_i \ov{\chi}_{i^*} \ , 
\nl
Z({\bf T}'_{c}) &=& 
\left ( \s_o \overline{\t_v} +  \t_v \overline{\s_o} 
+\s_v \overline{\t_o}+ \t_o \overline{\s_v}  \right ) 
+ \sum_i \chi_i \ov{\chi}_{i^*} \ , 
\nl 
Z({\bf T}'_{ca}) &=& 
\left ( \s_o \overline{\t_v} +  \t_v \overline{\s_o} 
 + |\s_v|^2 + |\t_o|^2 \right ) + \sum_i \chi_i \ov{\chi}_{i^*} \ .   
\label{z-stetra}
\ea
In these expressions, the index $i$ runs over all the fields
not explicitly present in the first parenthesis,
the prime indicates the presence of discrete torsion and
the subscript $a$ stands for the previous automorphism. 
The diagonal modular invariant $Z({\bf T}_d)$ similarly generates  
three other partition functions, that are denoted by  
$Z({\bf T}_{da})$, $Z({\bf T}_{d}')$ and $Z({\bf T}_{da}')$; 
they differ from the expressions (\ref{z-stetra}) by the substitution 
$\sum_i \chi_i \ov{\chi}_{i^*} \rightarrow \sum_i \chi_i \ov{\chi}_{i}$.

The boundary states for the model  ${\bf T}'_{c}$ with
discrete torsion are given by the action of the simple current
$\x_0$ on the Cardy boundaries of the ${\bf T}_{c}$ model. 
The expected number of 31 boundary states is reproduced, because the 
simple current $\x_0$ acting on the tetrahedron chiral fields 
forms 13 length-two orbits and 9 fixed points.

The boundary states for the ${\bf T}_{da}$ model can be obtained
by using the simple-current extension from the ${\bf O}$ model,
as already found in the bosonic case (Section 4); 
this will be discussed further 
below. The ${\bf T}'_{da}$ model can also be analyzed
by combining the previous two approaches.
Unfortunately, the boundaries for the other four cases 
${\bf T}_{ca}$, ${\bf T}'_{ca}$, ${\bf T}_{d}$ and ${\bf T}'_{d}$ 
do not seem to follow from simple-current constructions.

{\bf O}. We now turn to the discussion of the supersymmetric
octahedron model. This possesses 49 primary fields, 
30 belonging to the NS sector and 19 to the R sector, that are
all self-conjugate (see Appendix \ref{stoi-ch} for the character list
and the $S$ matrix).
There are three simple currents, $u_-,v_+,v_-$,  that form 
the group $\mathbb{Z}_2 \times \mathbb{Z}_2$ and can be used to build
several modular invariants. 
The current $u_-$ has integer spin and give the extension of 
the octahedron to the tetrahedron; the current $v_+$
has half-integer spin and contains the supercurrent:
the corresponding modular invariant 
coincide with the orbifold by $(-1)^{F_s}$, namely $\OO'$. 
Finally, the current $v_-$ has half-integer spin
and yields a new automorphism modular invariant,
that is named $Z(\widetilde{\OO})$.
The expressions of these modular invariants are, besides the diagonal
one:
\ba
Z({\bf O}') &=& \left ( \m_{0v} \overline{\m_{1o}} + 
\m_{0o} \overline{\m_{1v}}  
+ \s_{o+} \ov{\t}_{v+} +  \s_{o-} \ov{\t}_{v-} + c.c. \right ) + 
\sum_i |\chi_i |^2\ , 
\\
Z({\bf \widetilde{O}}) &=&  \left ( \a_0 \ov{\b}_2 + \a_1 \ov{\b}_3
+ \a_2 \ov{\b}_0 + \a_3 \ov{\b}_1 + \r_+\ov{\r}_- 
+ \s_{o+} \ov{\t}_{v-} +  \s_{o-} \ov{\t}_{v+} + c.c. \right ) 
\nonumber \\
&+& \sum_i |\chi_i |^2 \ , 
\\
Z({\bf \widetilde{O}}') &=& 
( \a_0 \ov{\b}_2 + \a_1 \ov{\b}_3 
+ \a_2 \ov{\b}_0 + \a_3 \ov{\b}_1 + \r_+\ov{\r}_- 
+ \m_{0v} \ov{\m}_{1o} + \m_{0o} \ov{\m}_{1v}  
\nonumber \\
&+& \s_{o+} \ov{\s}_{o-} +  \t_{v+} \ov{\t}_{v-} + c.c. ) + 
\sum_i |\chi_i |^2\ , 
\ea
where again the sums over $|\chi_i|^2$ contain all the fields not
explicitly written in the expressions.
Let us discuss these partition functions in turn.

${\bf O}'$. In this model there are $41$ 
Ishibashi states and the simple current
$v_+$ has precisely 19 length-two orbits and 11 fixed points, corresponding
to the chiral fields $\r_{\pm}$, $\r$, $osv$, $\L_i$, and $\g_i$.

${\bf \widetilde{O}}$. 
There are $35$ Ishibashi states: under the action of  
$v_-$, the octahedron fields form 21 orbits of length two 
and seven fixed points,
that split developing charges with respect to the
seven twisted characters $\r$, $osv$, $\L_i$, $i=0,1,2$, and $\m_{\a s}$,
$\a=0,1$. 
The boundary coefficients $R_{ai}$ (multiplied by $48 \sqrt{2}$)
of the resulting 14 boundaries w.r.t. to the Ishibashi states of 
the twisted fields (see Eqs.(\ref{frac-b}),(\ref{fix-s})), are the following:

$$
\begin{array}{c|ccccccc}
&\cdots & \r & osv & \L_0 &  \L_1 & \L_2 & \m_{\b s} 
\\ \hline
\cdots &\cdots &\cdots &\cdots &\cdots &\cdots &\cdots &\cdots
\\
|\r,\pm \ra &\cdots & 0 & 0 & \pm 16 \sqrt{3} & \pm 16 \sqrt{3} 
& \pm 16 \sqrt{3} & 0 
\\
|osv,\pm \ra & \cdots & 0 & 0& 0 &  0& 0 &\pm 24 \sqrt{2}(-1)^\b 
\\
|\L_0 ,\pm\ra&\cdots & \pm 16 \sqrt{3} & 0 & \mp 32 s_2 &\mp 32 s_4 
&\pm 32 s_1 & 0 
\\
|\L_1 ,\pm\ra&\cdots & \pm 16 \sqrt{3} & 0 & \mp 32 s_4 &\pm 32 s_1 
&\mp 32 s_2 & 0 
\\
|\L_2 ,\pm\ra&\cdots & \pm 16 \sqrt{3} & 0 & \pm 32 s_1 &\mp 32 s_2 
&\mp 32 s_4 & 0 
\\
|\m_{\a s}, \pm \ra&\cdots & 0 & \pm 24 \sqrt{2}(-1)^\a & 0 &0 &0 & \pm 24 
\end{array}
$$ 

where $s_1 =  \sin(\p/9)$, $s_2 =  \sin(2 \p/9)$ and $s_4 =  \sin(4 \p/9)$. 

${\bf \widetilde{O}'}$. There are $31$ Ishibashi states:
the corresponding boundaries can be obtained by acting on the boundaries
of the $\OO$ models with the full simple-current group $\Z_2 \times\Z_2$.

The boundary states for the ${\bf T}_{da}$ model are obtained
from the ${\bf O}$ model, as follows.
The $\mathbb{Z}_2$ symmetry ${\cal R}$ relating
 the tetrahedron and octahedron models,  
${\bf T}/{\cal R} = {\bf O}$, induces the automorphism of 
the fusion rules made by the charge-conjugation composed with
the exchange $\s_v \leftrightarrow \t_o$.
The 9 boundary states for the  $Z({\bf T}_{da})$ modular invariant
are obtained from the 18 boundary states of the octahedron
corresponding to the ${\cal R}$-twisted sector, namely 
$\{\a_k,\b_k,\g_k,\m_{\a I}\}$, with $k=0,\dots,3$, $\a=0,1$ and
$I=o,v,s$.
They are coupled by the simple current $u_-$ according to:
\be
(\a_0,\a_1,\b_0,\b_1,\g_0,\g_1,\m_{0s},\m_{0o},\m_{0v})
\mapsto
(\a_2,\a_3,\b_2,\b_3,\g_2,\g_3,\m_{1s},\m_{1o},\m_{1v})\ .
\ee 
Extending the ${\cal R}$-twisted sectors of the octahedron, 
we then obtain the boundary states for the ${\bf T}_{da}$ model. 
In a similar way one can obtain  
the boundary states for the ${\bf T}'_{da}$ model. 


\section{Conclusions}

In this paper, we have shown a number of interesting features
of boundary conformal field theories on orbifold spaces. 
The orbifold and simple-current relations between different theories
can be extended to mappings for boundary states;
these yield complete sets of boundaries for 
non-charge-conjugation modular invariants, furnish examples of
symmetry-breaking boundary conditions 
and can be visualized geometrically.

The non-abelian $\TT-\OO-\II$ orbifold models at $c=1$ and $3/2$ 
present some interesting features.
The origin and properties of fractional Dirichlet-like branes 
are well understood, while the geometrical interpretation of the
branes associated to the twisted sectors is not complete,
lacking a clear picture of the $\TT-\OO-\II$ orbifold spaces.
The analysis of supersymmetric models could also be developed;
in particular, the study of the modular covariance conditions 
in the R-R sector, that
relate the Witten index to the Ramond charges \cite{nepo}.

In conclusion, we hope that the models analyzed in this paper
will provide a useful playground for future studies of D-branes.

\vskip 24pt
{\large \bf Acknowledgments} 

We would like to thank A. Sagnotti, B. Schellekens, 
C. Schweigert, P. Valtancoli and J.-B. Zuber for interesting discussions.
A.C. thanks the Theory Group at CERN for hospitality.
G.D. acknowledges the support by CNRS, France, LPTHE,
UMR 7589.

\newpage
\appendix


\section{Character tables of the ${\bf T}$-${\bf O}$-${\bf I}$ groups}
\label{toi-tab}
{\sc
$$
\begin{array}{l|rr|rrrr|r}
|C_a| & 1 & 1 & 4 & 4 & 4 & 4 & 6
\\
i_a & I & -I & T & T^{-1}& -T^{-1}& -T & S 
\\ \hline
u_0 & 1 & 1 & 1 & 1 & 1 & 1 & 1
\\
u_1 & 1 & 1 &\w  &\ov{\w} & \w  &\ov{\w}  & 1
\\
u_2 & 1 & 1 & \ov{\w} &\w  &\ov{\w}  & \w & 1
\\
j  & 3 & 3 & 0 &  0& 0 & 0 & -1
\\
\f_0 &  2 & -2 & -1 & -1 & 1 & 1 & 0
\\
\f_1 & 2 & -2 & -\ov{\w} &-\w  &\ov{\w}  &\w  & 0
\\
\f_2  & 2 & -2 & -\w & -\ov{\w} & \w &\ov{\w}  & 0
\end{array}
$$
$$ $$
$$
\begin{array}{l|rr|rr|r|rrr}
|C_a| & 1 & 1 & 8 & 8 & 12 & 6 & 6 & 6
\\
i_a & I & -I & T & -T^{-1}& E & S & F& -F^T 
\\ \hline
u_+ & 1 & 1 & 1 & 1 & 1 & 1 & 1 & 1
\\
u_- & 1 & 1 & 1   &1 & -1 & 1  & -1 &-1
\\
u_f & 2 & 2 & -1  &-1  & 0 & 2 &0 &0
\\
j_+ & 3 & 3 & 0 & 0& 1 & -1 & -1 & -1
\\
j_- & 3 & 3 & 0 & 0& -1 & -1 & 1 & 1
\\
\f_+& 2 & -2 & -1 & 1 & 0 & 0 & \sqrt{2} &-\sqrt{2}
\\
\f_-& 2 & -2 &  -1 &1 &  0& 0 &-\sqrt{2} &\sqrt{2}
\\
\f_f& 4 & -4 & 1 & -1& 0 & 0 & 0 & 0
\end{array}
$$
$$ $$
$$
\begin{array}{l|cc|cccc|c|cc}
|C_a| & 1 & 1 & 12 & 12 & 12 & 12 & 30 & 20 & 20
\\
i_a & I& -I& T& T^2& -T^{-1}& T^{-2}& 2 E& 2H& 2H^{-1}
\\ \hline
u_0 & 1 & 1 & 1 & 1 & 1 & 1 & 1 & 1 & 1
\\
u_1 & 3 & 3 & A & \widetilde{A} & A & \widetilde{A}  & -1 & 0 & 0
\\
u_2 & 3 & 3 & \widetilde{A} & A & \widetilde{A} & A  & -1 & 0 & 0
\\
u_3 & 4 & 4 & -1 & -1& -1 & -1 & 0 & 1 & 1
\\
u_4 & 5 & 5 & 0 & 0& 0 & 0 & 1 & -1 &-1
\\
\f_1& 2 & -2 & -A & -\widetilde{A} & A & \widetilde{A} & 0 & -1 & 1
\\
\f_2& 2 & -2 & -\widetilde{A} & -A & \widetilde{A} & A & 0 & -1 & 1
\\
\f_3& 4 & -4 &  -1 & -1 &  1& 1 &0 & 1 & -1
\\
\f_4& 6 & -6 & 1 & 1& -1 & -1 & 0 & 0 & 0
\end{array}
$$
}

These are the character tables of the groups: $SL_2(\Z_3)$ (top), 
$GL_2(\Z_3)$ (center) and $SL_2(\Z_5)$ (bottom).
We used the notations: $\w=\exp(2i\pi/3)$
$A=(1+\sqrt{5})/2$ and $\widetilde{A}=(1-\sqrt{5})/2$.
The representative elements $i_a$ of each conjugacy class are:

{\sc
$
I= \left( \begin{array}{cc} 1& 0\\ 0& 1\end{array} \right) , \ 
T= \left(  \begin{array}{cc}1 & 1\\0 & 1\end{array}\right) , \ 
S= \left( \begin{array}{cc} 0& -1\\ 1& 0\end{array}\right) ,\ 
E= \left( \begin{array}{cc} 1& 0\\ 0& -1\end{array}\right) , \ 
F= \left( \begin{array}{cc} 1& -1\\ 1& 1\end{array}\right) , \  
H= \left( \begin{array}{cc} 1& 2\\ 1& 1\end{array}\right)\ .
$
}

\section{Amplitudes of non-orientable surfaces}
\label{klein}

We complete the discussion of the boundary states 
for the rational CFTs at $c=1$ (Sections 3 and 4) by
considering the amplitudes on the Klein bottle
and the M\"obius strip \cite{carg}. These amplitudes
project the closed and open spectra onto  
states invariant under the world-sheet parity operation $\W$. 
The Klein bottle can be written as:
\be
K = \frac{1}{2} \sum_i K^i\ \chi_i \ ,
\ee
where the coefficients $K^i$ are constrained by the requirement
of integrality and positivity of the partition function $Z/2 + K$. 
Upon $S$ modular transformation, this amplitude describes
the propagation of states in the closed sector between two crosscap
states:
\be
\widetilde{K} = \sum_i \G_i^2\ \widetilde{\chi}_i \ ,
\ee
where the $\G_i$ are the so-called crosscap coefficients.

The open spectrum is described by the annulus and M\"obius amplitudes;
the first one is:
\be
A=\frac{1}{2}\ \sum_{i,a,b}\ n^a\ n^b \ A^i_{ab}\ \chi_i\ ,
\ee
where we now sum over all boundaries with multiplicities $n^a,n^b$.
The M\"obius amplitude reads:
\be
M = \pm \frac{1}{2} \sum_{i,a} n^a\ M_a^{\ i}\ \widehat{\chi}_i  \ ,
\ee
where the hatted characters \cite{carg} are defined as 
$\widehat{\chi} = T^{-1/2} \chi $ and
the M\"obius coefficients $M_a^{\ i}$ are again
constrained by the requirement of integrality and positivity
of the partition function $A+M$.
The transverse M\"obius amplitude is obtained through the modular 
transformation $P= T^{1/2}ST^2ST^{1/2}$, and describes the propagation 
of closed string states between a boundary and a crosscap:
\be
\widetilde{ M } = \pm \sum_{i,a} \G_i \ B_{ai} \ 
\widetilde{\widehat{\chi}}_i \ .
\ee

In general, given a bulk conformal field theory one has several choices
for the Klein bottle amplitude \cite{carg}\cite{su2}. 
These different Klein bottle projections
correspond to acting on the closed and open spectra with a combination
of $\W$ and some other involution of the theory, that can often be described
by simple-current techniques \cite{s-klein}\cite{geom-orien}.

The Cardy solution for the annulus boundaries of 
the charge-conjugation modular invariant has been extended
to the Klein and M\"obius coefficients  \cite{su2}; it can be presented
in a nice way by introducing the tensor:
\be
Y_{ij}{}^k = \sum_l \frac{S_{il}P_{jl}P^{\dagger}_{kl}}{S_{0l}} \ .
\label{yassen}
\ee
The ansatz for the crosscap coefficients is:
\be
\G_i = \frac{P_{0i}}{\sqrt{S_{0i}}} \ ,
\label{kb1}
\ee
from which one can derive,
\be
K_i = Y_{i00}\ , \qquad\qquad M_{ai} = Y_{ai0} \ .
\ee 
As shown in \cite{s-klein}, one can define a modified Klein bottle 
projection whenever the model contains a simple current $J$. The 
corresponding crosscap coefficients are:
\be
\G_i = \frac{P_{J,i}}{\sqrt{S_{J,i}}} \ ,
\label{kb2}
\ee
from which one can derive,
\be
K_i = Y_{i,J}{}^{J}\ , \qquad\qquad M_{a}{}^{i} = Y_{J^*(a),J}{}^{i} \ .
\ee 
Moreover, this ansatz has been extend to simple-current modular invariants
in Ref.\cite{bcs}. 

Let us now discuss the compactified boson at $c=1$.
We have two natural projections: $\W$ and 
$\W \d_2$, where $\d_2$ is the half-radius shift operator  
$X \mapsto X + \p R$.  
With the charge-conjugation modular invariant,
only the representations $\chi_0$ and $\chi_k$ can appear in the Klein bottle
which reads, in the two cases: 
\be
K_{r} = \frac{1}{2}( \chi_0 +  \chi_k ) \ ,
\hspace{1cm} {K}_{c} = \frac{1}{2}( \chi_0 +  (-1)^k \chi_k ) \ . 
\ee
The two projections are really distinct for $k$ odd. 

We can then construct the annulus and M\"obius amplitudes. 
For example, their explicit expressions for $k=3$ are:
\ba
A_r &=& 
\frac{1}{2}  \left[ \chi_0 \left ( n_0^2+n_3^2+2n\ov{n}+2m\ov{m} \right ) + 
\chi_1 \left ( 2n_0n+2m\ov{n}+2n_3\ov{m} \right )  \right.
\nl 
&&+ \chi_2 \left ( n^2+\ov{m}^2+2n_0m+2n_3\ov{n} \right )
 +  \chi_3 \left ( 2n_0n_3+2nm+2\ov{n}\ov{m} \right ) 
\nl 
&& \left. +\chi_{-2} \left ( m^2+\ov{n}^2+2n n_3+2n_0\ov{m} \right ) +
\chi_{-1} \left ( 2n_0\ov{n}+2n\ov{m}+2n_3 m \right ) \right] \ , 
\nl 
M_r &=& \pm \frac{1}{2} 
\left [ \hat{\chi}_0(n_0-n_3)+\hat{\chi}_2(n-\ov{m})
+\hat{\chi}_{-2}(\ov{n}-m) \right ] \ ,
\ea
and
\ba
A_c &=& \frac{1}{2}  \left[ \chi_0 
\left ( 2l\ov{l}+2n\ov{n}+2m\ov{m} \right ) + 
\chi_1 \left ( 2l\ov{m}+2n\ov{l}+m^2 + \ov{n}^2 \right ) \right.
\nonumber \\
&&+ \chi_2 \left ( 2\ov{l}\ov{n}+2n\ov{m}+2m l \right ) 
+ \chi_3 \left ( l^2+\ov{l}^2+2m n + 2 \ov{m}\ov{n} \right ) 
\nonumber \\
&& \left. +\chi_{-2} \left ( 2ln+2m\ov{n}+2\ov{l}\ov{m} \right ) +
\chi_{-1} \left ( 2\ov{l}m+2 l \ov{n}+n^2+\ov{m}^2 \right )  \right] \ , 
\nl
M_c &=& \pm \frac{1}{2} 
\left [ \hat{\chi}_1(m+\ov{n})+\hat{\chi}_3(l+\ov{l})+\hat{\chi}_{-1}(n+\ov{m}) \right ] \ .
\label{od-circle}
\ea
In Eq. (\ref{od-circle}), boundaries that are not self-conjugate 
($A^0_{aa} = 0$) carry a pair of complex charges, e.g. $l,\bar{l}$.  

It is clear from Eq.(\ref{od-circle}) that different $\W$ projections 
lead to different boundary conjugation properties. 
When a geometric interpretation is available \cite{geom-orien}, 
the boundary conjugation properties simply reflect the action of
$\W$ on the submanifolds wrapped by the brane world-volume. 
In the simple case analyzed before, 
$\W$ maps $X \mapsto - X$ and therefore all the 
branes sitting at opposite position form conjugate pairs 
and carry complex charges, 
except for the branes sitting at $X=0$ and $X=\pi R$ 
which are fixed under $\W$ and carry real charges. 
In a similar way, $\W \d_2$ maps $X \mapsto -X + \pi R$ and
the fixed branes are those sitting at $X= \pm \pi R/2$.
From the point of view of the rational CFT, the second 
projection gives rise, for $k$ odd, to an annulus amplitude involving only
complex charges: the two fixed branes are missing, because they
do not correspond to symmetry-preserving boundary conditions.

Let us consider now the diagonal modular invariant.  
In this case all the characters can appear in the Klein 
bottle and the two projections read:
\be
{K}_1 = \frac{1}{2} \sum_{i=0}^{2k-1} \chi_i \ , 
\hspace{1cm} {K}_2 = \frac{1}{2} \sum_{i=0}^{2k-1} (-1)^i \chi_i \ .
\ee
Recall that the annulus only contains the two boundaries 
in (\ref{diag-bound}), obtained by the orbifold construction. 
For example, for $k=3$, we find:
\ba
A_1 &=& n \ov{n} (\chi_0 + \chi_2 + \chi_{-2})
+ \frac{1}{2} \left ( n^2+\ov{n}^2 \right ) (\chi_1+\chi_3+\chi_{-1}) \ , 
\nonumber \\
M_1 &=& \pm \frac{1}{2} \left ( n + \ov{n} \right )  (\hat{\chi}_1
+\hat{\chi}_3+\hat{\chi}_{-1}) \ , \nonumber \\
A_2 &=& \frac{1}{2} \left ( n_+^2+n_-^2 \right ) 
(\hat{\chi}_0 + \hat{\chi}_2 +\hat{\chi}_{-2})
+ n_+n_- (\chi_1+\chi_3+\chi_{-1}) \ , 
\nonumber \\
M_2 &=& 
\pm \frac{1}{2} \left ( n_+ +  n_- \right ) (\chi_0 - \chi_2 - \chi_{-2})   \ .
\ea
It is easy to verify that these amplitudes are consistent 
in the transverse channel.

A similar analysis can be performed for the orbifold line, 
in particular one can see that all the charges are real except for  
those corresponding to the fractional branes at the two fixed points,
which can be real or complex depending on the action of $\W$ 
on the twisted sectors.

We now describe in some detail the Klein bottle projection for the
$\TT$ model.
For the charge-conjugation modular invariant the standard
Klein bottle projection is:
\be
{K} = \frac{1}{2}(u_0 + \f_0 + j + \s + \t) \ .
\ee
For the diagonal modular invariant, it is simply given by:
\be
K_d = \frac{1}{2} \sum_{i=1}^{21} \chi_i \ .
\label{ktdiag}
\ee
It is interesting to notice that 
the crosscap coefficients for the diagonal case can be obtained from the 
crosscap coefficients of the $\OO$ model by acting with the simple current
$u_-$, as done for the annulus coefficients in Section 4.
The crosscap coefficients for the diagonal $\TT$ model are thus found to be:
\be
\G_i = \frac{P_{0,i}-P_{u_-, i}}{\sqrt{2S_{0,i}}} 
= \frac{1}{2^{1/4}}
\left(\sqrt{6},\sqrt{2},0,1,1 \right)\ ,
\ee
where the P and $S$ matrices are relative to 
the $\OO$ model, and the five entries in the vector refer to the tetrahedron 
primaries $(u_0,j,\f_0,\s,\t)$, respectively. 
In the direct channel, these crosscap coefficients
give the amplitude (\ref{ktdiag}).

The other models at $c=1$ and at $c=3/2$, with various choices 
for the Klein bottle, can be discussed along similar lines,
using the Eqs. (\ref{kb1},\ref{kb2}).

\section{Chiral data of $\TT-\OO-\II$ models}
\label{toi-ch}
\subsection{Tetrahedron}
The fusion rules of the theory, in the field basis given in 
Section 4, are the following $(i,j=0,1,2 \ {\rm mod} \ 3)$:
\ba
u_i u_j &=& u_{i+j} \ ,  \hspace{1cm} \f_i \f_j = u_{i+j} + j \ , 
\hspace{1cm}  
jj = \sum_{i=0}^2 u_i +2j \ , \nl 
\s \s = \t \t &=&  \sum_{i=0}^2 u_i +j + \sum_{i=0}^2 \f_i +2 \s + 2 \t \ , 
\hspace{0.5cm}
\s \t = 2j + \sum_{i=0}^2 \f_i +2 \s + 2 \t \ , \nl
\w^{\pm}_i \w^{\pm}_j &=& \w^{\mp}_{1-i-j} + \sum_{k=0}^2 \th^{\mp}_k \ , 
\hspace{2.4cm} 
\w^+_i \w^-_j = u_{j-i}+j+\s+\t \ , \nl
\w_i^{\pm}\th^{\pm}_j &=&  \sum_{k=0}^2 \w^{\mp}_k + \th^{\mp}_{i-j} \ , 
\hspace{2.9cm}  
\w_i^+\th_j^- = \s + \t + \sum_{k \ne 2-i-j}^2 \f_k \ , \nl
\th^{\pm}_i \th^{\pm}_j &=&  \sum_{k=0}^2 \th^{\mp}_k + \w^{\mp}_{i+j} \ , 
\hspace{3.0cm} 
\th^{+}_i \th^{-}_j = j + \s + \t + u_{i-j} \ . 
\label{tetra-fus}
\ea

Hereafter, we report the annulus amplitudes for the theory with
diagonal modular invariant. There are 5 boundary states,
described in Section 4.2 (Eqs.(\ref{td-bs}),(\ref{td-s})); 
the corresponding annulus coefficients
$A^n_{ab}$, with $a,b=1,\dots,5$, 
can be written as $5\times 5$ matrices; the index $n$ runs over the
21 chiral sectors of the theory, ordered as in the $S$-matrix (\ref{tetra-s}),
$\{[\f_n] \ |\ n=1,\dots,21\}\equiv
\{ u_i,j,\f_i,\s,\t,\w^+_i,\w^-_i,\th^+_i,\th^-_i|\ i=0,1,2\}$.
The matrices are: $A^n= {\bf 1}_5$, $n=1,2,3$, and 
\be
{\sc
\begin{array}{ll}
A^4= \left( \begin{array}{ccccc}
1&2 & 0&0 &0 \\
2& 1& 0& 0& 0\\
0& 0& 1& 2& 0\\
0& 0& 2&1 &0 \\
0& 0& 0& 0&3 
\end{array} \right), 
& 
A^m= \left( \begin{array}{ccccc}
0 & 0 & 1& 1&0\\
0 & 0 & 1& 1&0\\
1& 1& 0& 0&0\\
1& 1& 0& 0&0\\
0& 0& 0&0 &2
\end{array} \right), 
\ m=5,6,7,
\\
A^8= \left( \begin{array}{ccccc}
1& 0& 1& 0& 2\\
0&1 &0 & 1&2 \\
1&0 &0 &1 & 2\\
0& 1& 1&0 & 2\\
2& 2& 2& 2&2
\end{array} \right)\ , 
& 
A^9= \left( \begin{array}{ccccc}
0&1 &0 & 1&2 \\
1& 0& 1& 0& 2\\
0& 1& 1&0 & 2\\
1&0 &0 &1 & 2\\
2& 2& 2& 2&2
\end{array}\right)
\\  
A^l= \left( \begin{array}{ccccc}
0& 0& 1& 1&1\\
0& 0& 1& 1&1\\
1&1 & 0& 0&1\\
1&1 & 0& 0&1\\
1& 1& 1& 1&2
\end{array}\right),
\ l=10,\dots,15,
&
A^k= \left( \begin{array}{ccccc}
1& 1& 0& 0&1\\
1& 1& 0& 0&1\\
0& 0& 1&1 &1\\
0& 0& 1&1 &1\\
1 & 1& 1& 1 & 2
\end{array}\right),
\ k=16,\dots,21.
\end{array}}
\label{tetra-an}
\ee
One can check that they give a representation of the fusion rules
(\ref{tetra-fus}).


%
\subsection{Octahedron}
Untwisted sector:
\be
\begin{array}{llll}
u_+ &= \frac{1}{24} \th\!\left[^{0}_0 \right] + 
\frac{1}{3} \th\!\left[^{\ 0}_{1/3} \right]
+ \frac{3}{8} \th\!\left[^{\ 0}_{1/2} \right] 
+ \frac{1}{4} \th\!\left[^{\ 0}_{1/4} \right]\ , 
& &h = 0 \ ,\\
u_- &=  \frac{1}{24} \th\!\left[^0_0 \right] + 
\frac{1}{3} \th\!\left[^{\ 0}_{1/3} \right]
- \frac{1}{8} \th\!\left[^{\ 0}_{1/2} \right] 
- \frac{1}{4} \th\!\left[^{\ 0}_{1/4} \right] \ ,
&  &h=9 \ ,\\
u_f &= \frac{1}{12} \th\!\left[^0_0 \right] 
- \frac{1}{3} \th\!\left[^{\ 0}_{1/3} \right]
+ \frac{1}{4} \th\!\left[^{\ 0}_{1/2} \right]  \ ,
&  &h=4 \ ,\\
j_+ &=  \frac{1}{8} \th\!\left[^0_0 \right] 
+ \frac{1}{8} \th\!\left[^{\ 0}_{1/2} \right]
- \frac{1}{4} \th\!\left[^{\ 0}_{1/4} \right]  \ ,
& &h=4 \ ,\\
j_- &= \frac{1}{8} \th\!\left[^0_0 \right] 
- \frac{3}{8} \th\!\left[^{\ 0}_{1/2} \right]
+ \frac{1}{4} \th\!\left[^{\ 0}_{1/4} \right]  \ ,
& &h=1 \ ,\\
\f_+ &= \frac{1}{12} \th\!\left[^{1/2}_{\ 0} \right] 
- \frac{\ov{\w}}{3} \th\!\left[^{1/2}_{1/3} \right]
+ \frac{1}{4} \th\!\left[^{1/2}_{1/4} \right]
+ \frac{1}{4} \th\!\left[^{1/2}_{3/4} \right]   \ ,
& &h=\frac{1}{4} \ ,\\
\f_- &= \frac{1}{12} \th\!\left[^{1/2}_{\ 0} \right] 
- \frac{\ov{\w}}{3} \th\!\left[^{1/2}_{1/3} \right]
- \frac{1}{4} \th\!\left[^{1/2}_{1/4} \right]
- \frac{1}{4} \th\!\left[^{1/2}_{3/4} \right]   \ , 
&&h=\frac{25}{4} \ ,\\
\f_f &= \frac{1}{6} \th\!\left[^{1/2}_{\ 0} \right] 
+ \frac{\ov{\w}}{3} \th\!\left[^{1/2}_{1/3} \right]  \ ,
& &h=\frac{9}{4}. 
\end{array}
\label{oct1}
\ee
$\mathbb{Z}_2$-twisted sector $(i=0,1)$:
\be
\begin{array}{llll}
\m_i &= \frac{1}{2} \th\!\left[^{1/4}_{\ 0} \right] + 
\frac{(-1)^i}{2} \th\!\left[^{1/4}_{1/2} \right] \ ,
&&h=\frac{1}{16} \ , \frac{9}{16} \ . 
\end{array}
\label{oct2}
\ee
$\mathbb{Z}_3$-twisted sector ($ i = 0, 1, 2$):
\be
\begin{array}{llll}
\w_i &=  \frac{1}{3} \th\!\left[^{1/3}_{\ 0} \right] 
+\frac{\w^i}{3} \th\!\left[^{1/3}_{1/3} \right]
+\frac{\ov{\w}^i}{3} \th\!\left[^{1/3}_{2/3} \right] \ ,
&&h=\frac{1}{9}, \frac{4}{9}, \frac{16}{9}\ ,
 \\
\th_i &=  \frac{1}{3} \th\!\left[^{1/6}_{\ 0} \right] 
+\frac{\w^i}{3} \th\!\left[^{1/6}_{1/3} \right]
+\frac{\ov{\w}^i}{3} \th\!\left[^{1/6}_{2/3} \right] \ ,
&&h=\frac{1}{36}, \frac{25}{36}, \frac{49}{36}\ .
 \\
\end{array}
\label{oct3}
\ee
$\mathbb{Z}_4$-twisted sector ($ k = 0, 1, 2, 3$):
\be
\begin{array}{llll}
\a_k &= \frac{1}{4} \th\!\left[^{1/8}_{\ 0} \right] 
+\frac{i^k}{4} \th\!\left[^{1/8}_{1/4} \right]
+\frac{\left (-1\right)^k}{4} \th\!\left[^{1/8}_{1/2} \right] 
+\frac{\left (-i\right)^k}{4} \th\!\left[^{1/8}_{3/4} \right] \ ,
&&h=\frac{1}{64}, \frac{49}{64}, \frac{225}{64}, \frac{81}{64} \ , 
\\ 
\b_k &=  \frac{1}{4} \th\!\left[^{3/8}_{\ 0} \right] 
+\frac{i^k}{4} \th\!\left[^{3/8}_{1/4} \right]
+\frac{\left (-1\right)^k}{4} \th\!\left[^{3/8}_{1/2} \right] 
+\frac{\left (-i\right)^k}{4} \th\!\left[^{3/8}_{3/4} \right] \ ,
&&h=\frac{9}{64}, \frac{25}{64}, \frac{169}{64}, \frac{121}{64} \ ,
\\
\s_{\pm} &=\frac{1}{4} \th\!\left[^{1/4}_{\ 0} \right] 
+\frac{1}{4} \th\!\left[^{1/4}_{1/2} \right]
\pm \frac{1}{4} \th\!\left[^{1/4}_{1/4} \right] 
\pm \frac{1}{4} \th\!\left[^{1/4}_{3/4} \right]\ ,
&&h=\frac{1}{16}  \ , \frac{49}{16}  \ , \\ 
\t_{\pm} &=\frac{1}{4} \th\!\left[^{1/4}_{\ 0} \right] 
-\frac{1}{4} \th\!\left[^{1/4}_{1/2} \right]
\pm \frac{i}{4} \th\!\left[^{1/4}_{1/4} \right] 
\mp \frac{i}{4} \th\!\left[^{1/4}_{3/4} \right]\ ,
&&h=\frac{9}{16} \ ,  \frac{25}{16} \ .  
\end{array}
\label{oct4}
\ee


\subsection{Icosahedron}
Untwisted sector:
\be
\begin{array}{llll}
u_0 &= \frac{1}{60} \th\!\left[^0_0 \right] + 
\frac{1}{3} \th\!\left[^{\ 0}_{1/3} \right]
+ \frac{1}{5} \th\!\left[^{\ 0}_{1/5} \right] 
+ \frac{1}{5} \th\!\left[^{\ 0}_{2/5} \right]
+ \frac{1}{4} \th\!\left[^{\ 0}_{1/2} \right] \ , 
&&h = 0\ , \\
u_1 &= \frac{1}{20} \th\!\left[^0_0 \right] 
+ \frac{1+\sqrt{5}}{10} \th\!\left[^{\ 0}_{1/5} \right] 
+ \frac{1-\sqrt{5}}{10} \th\!\left[^{\ 0}_{2/5} \right]
- \frac{1}{4} \th\!\left[^{\ 0}_{1/2} \right] \ , 
&&h =1\ , \\
u_2 &= \frac{1}{20} \th\!\left[^0_0 \right] 
+ \frac{1-\sqrt{5}}{10} \th\!\left[^{\ 0}_{1/5} \right] 
+ \frac{1+\sqrt{5}}{10} \th\!\left[^{\ 0}_{2/5} \right]
- \frac{1}{4} \th\!\left[^{\ 0}_{1/2} \right] \  
&&h =9 \ ,\\
u_3 &= \frac{1}{15} \th\!\left[^0_0 \right] + 
\frac{1}{3} \th\!\left[^{\ 0}_{1/3} \right]
- \frac{1}{5} \th\!\left[^{\ 0}_{1/5} \right] 
- \frac{1}{5} \th\!\left[^{\ 0}_{2/5} \right] \  ,
&&h = 9 \ ,\\
u_4 &= \frac{1}{12} \th\!\left[^0_0 \right] - 
\frac{1}{3} \th\!\left[^{\ 0}_{1/3} \right]
+ \frac{1}{4} \th\!\left[^{\ 0}_{1/2} \right] \  ,
&&h = 4\ , \\
\f_1 &= \frac{1}{30} \th\!\left[^{1/2}_0 \right] 
- \frac{1+\sqrt{5}}{10} \ov{\z}^2 \th\!\left[^{1/2}_{1/5} \right] 
- \frac{1-\sqrt{5}}{10} \z \th\!\left[^{1/2}_{2/5} \right]
- \frac{\ov{\w}}{3} \th\!\left[^{1/2}_{1/3} \right] \ , 
&&h = \frac{1}{4} \ ,\\
\f_2 &= \frac{1}{30} \th\!\left[^{1/2}_0 \right] 
- \frac{1-\sqrt{5}}{10} \ov{\z}^2 \th\!\left[^{1/2}_{1/5} \right] 
- \frac{1+\sqrt{5}}{10} \z \th\!\left[^{1/2}_{2/5} \right]
- \frac{\ov{\w}}{3} \th\!\left[^{1/2}_{1/3} \right] \  ,
&&h = \frac{49}{4} \ ,\\
\f_3  &= \frac{1}{15} \th\!\left[^{1/2}_0 \right] 
- \frac{1}{5} \ov{\z}^2 \th\!\left[^{1/2}_{1/5} \right] 
- \frac{1}{5} \z \th\!\left[^{1/2}_{2/5} \right]
+ \frac{\ov{\w}}{3} \th\!\left[^{1/2}_{1/3} \right] \ , 
&&h = \frac{9}{4} \ ,\\
\f_4 &= \frac{1}{10} \th\!\left[^{1/2}_0 \right] 
+ \frac{1}{5} \ov{\z}^2 \th\!\left[^{1/2}_{1/5} \right] 
+ \frac{1}{5} \z \th\!\left[^{1/2}_{2/5} \right] \ ,
&&h = \frac{25}{4}\ .
\end{array}
\label{ico1}
\ee
$\mathbb{Z}_2$-twisted sector:
\be
\begin{array}{llll}
\s &= \frac{1}{2} \th\!\left[^{1/4}_{\ 0} \right] + 
\frac{1}{2} \th\!\left[^{1/4}_{1/2} \right] \ ,
&&h=\frac{1}{16}  \ ,\\
\t &= \frac{1}{2} \th\!\left[^{1/4}_{\ 0} \right] -
\frac{1}{2} \th\!\left[^{1/4}_{1/2} \right]  \ ,
&&h=\frac{9}{16}\ .
\end{array}
\label{ico2}
\ee
$\mathbb{Z}_3$-twisted sector ($ i = 0, 1, 2$):
\be
\begin{array}{llll}
\w_i &=  \frac{1}{3} \th\!\left[^{1/3}_{\ 0} \right] 
+\frac{\w^i}{3} \th\!\left[^{1/3}_{1/3} \right]
+\frac{\ov{\w}^i}{3} \th\!\left[^{1/3}_{2/3} \right] \ ,
&&h=\frac{1}{9}, \frac{4}{9}, \frac{16}{9}\ ,
 \\
\th_i &=  \frac{1}{3} \th\!\left[^{1/6}_{\ 0} \right] 
+\frac{\w^i}{3} \th\!\left[^{1/6}_{1/3} \right]
+\frac{\ov{\w}^i}{3} \th\!\left[^{1/6}_{2/3} \right] \ ,
&&h=\frac{1}{36}, \frac{25}{36}, \frac{49}{36}\ .
\end{array}
\label{ico3}
\ee
$\mathbb{Z}_5$-twisted sector ($ k = 0, 1, 2, 3, 4$):
\be
\begin{array}{llll}
\p_k &= \frac{1}{5} \th\!\left[^{1/5}_{\ 0} \right] 
+ \frac{\z^k}{5} \th\!\left[^{1/5}_{1/5} \right]
+ \frac{\z^{2k}}{5} \th\!\left[^{1/5}_{2/5} \right] 
&& \\
&\ \ + \frac{\ov{\z}^{2k}}{5} \th\!\left[^{1/5}_{3/5} \right]
+ \frac{\ov{\z}^k}{5} \th\!\left[^{1/5}_{4/5} \right] \ , 
&&h = \frac{1}{25}, \frac{16}{25}, \frac{81}{25}, 
\frac{121}{25}, \frac{36}{25}  \ , 
\\
\r_k &= \frac{1}{5}  \th\!\left[^{1/10}_{\ 0} \right] 
+ \frac{\z^k}{5} \th\!\left[^{1/10}_{1/5} \right]
+ \frac{\z^{2k}}{5} \th\!\left[^{1/10}_{2/5} \right] 
&& \\
&\ \ + \frac{\ov{\z}^{2k}}{5} \th\!\left[^{1/10}_{3/5} \right]
+ \frac{\ov{\z}^k}{5} \th\!\left[^{1/10}_{4/5} \right] \ ,
&&h = \frac{1}{100}, \frac{81}{100}, \frac{361}{100}, 
\frac{441}{100}, \frac{121}{100}  \ , 
\\
\l_k &= \frac{1}{5}  \th\!\left[^{2/5}_{\ 0} \right] 
+ \frac{\z^k}{5} \th\!\left[^{2/5}_{1/5} \right]
+ \frac{\z^{2k}}{5} \th\!\left[^{2/5}_{2/5} \right] 
&& \\
&\ \ + \frac{\ov{\z}^{2k}}{5} \th\!\left[^{2/5}_{3/5} \right]
+ \frac{\ov{\z}^k}{5} \th\!\left[^{2/5}_{4/5} \right] \ ,
&&h= \frac{4}{25}, \frac{9}{25}, \frac{64}{25}, \frac{144}{25}, 
\frac{49}{25} \ ,    
\\
\xi_k &= \frac{1}{5}  \th\!\left[^{3/10}_{\ 0} \right] 
+ \frac{\z^k}{5} \th\!\left[^{3/10}_{1/5} \right]
+ \frac{\z^{2k}}{5} \th\!\left[^{3/10}_{2/5} \right] 
&& \\
&\ \ + \frac{\ov{\z}^{2k}}{5} \th\!\left[^{3/10}_{3/5} \right]
+ \frac{\ov{\z}^k}{5} \th\!\left[^{3/10}_{4/5} \right] \ ,
&& h = \frac{9}{100}, \frac{49}{100}, \frac{289}{100}, 
\frac{529}{100}, \frac{169}{100}\ ,  
\end{array}
\label{ico4}
\ee
where $\z = \exp(2 i \pi/5)$.

The $S$ matrix ($\times 60\sqrt{2}$) in this basis is:
{\sc
$$
\begin{array}{c|ccccccccccccccccc}
& u_0 & u_1 & u_2 & u_3 & u_4 & \f_1 & \f_2 & \f_3 & 
\f_4 & \s & \t & \w_j & \th_j & \pi_l & \r_l & \l_l & \xi_l  
\\ \hline
u_0 & 1& 3& 3& 4& 5& 2& 2& 4& 6& 30& 30& 20&20& 
12& 12& 12& 12
\\
u_1 & 3& 9& 9& 12& 15& 6& 6& 12& 18& -30& -30& 0&0 &
\widetilde{g}& g & g &\widetilde{g}  
\\
u_2&  3& 9& 9& 12& 15& 6& 6& 12& 18& -30& -30& 0&0 &
g & \widetilde{g}& \widetilde{g} & g 
\\
u_3& 4& 12& 12& 16& 20& 8& 8& 16& 24& 0&0 & 20& 20 &
-12& -12& -12& -12 
\\
u_4& 5& 15& 15& 20& 25& 10& 10& 20& 30& 30& 30& -20&-20 &
0& 0& 0& 0 
\\
\f_1& 2& 6& 6& 8& 10& -4& -4& -8& -12& 0& 0& -20& 20 &
-g& \widetilde{g}& -\widetilde{g}& g 
\\
\f_2& 2& 6& 6& 8& 10& -4& -4& -8& -12& 0& 0& -20& 20 &
-\widetilde{g}& g & -g &\widetilde{g}  
\\
\f_3& 4& 12& 12& 16& 20& -8& -8& -16& -24& 0& 0& 20&-20 &
-12& 12& -12&12  
\\
\f_4& 6& 18& 18& 24& 30& -12& -12& -24& -36& 0& 0& 0& 0&
12& -12& 12& -12 
\\
\s& 30& -30& -30& 0& 30& 0& 0&0 &0 &30\sqrt{2} &-30\sqrt{2} &0 & 0&
0&0 &0 &0  
\\
\t& 30& -30& -30& 0& 30& 0& 0&0 &0 &-30\sqrt{2} &30\sqrt{2} &0 & 0&
0&0 &0 &0  
\\
\w_i& 20& 0& 0& 20& -20& -20& -20& 20& 0& 0& 0&a_{ij} & b_{ij}& 
0& 0& 0& 0 
\\
\th_i& 20& 0& 0& 20& -20& -20& -20& 20& 0& 0& 0&b_{ji} & d_{ij}& 
0& 0& 0& 0  
\\
\pi_k& 12& \widetilde{g}& g& -12& 0& -g& -\widetilde{g}& -12& 12& 0& 0& 0&0& 
P^1_{kl}& P^2_{kl}& P^3_{kl}& P^4_{kl} 
\\
\r_k&  12& g &\widetilde{g}& -12& 0& \widetilde{g}& g &  12& -12& 0& 0& 0&0& 
P^2_{lk}& R^1_{kl}& R^2_{kl}&R^3_{kl}  
\\
\l_k&  12& g &\widetilde{g}&  -12& 0&  -\widetilde{g}& - g & -12& 12& 0& 0& 0&0& 
P^3_{lk}& R^2_{lk}& L^1_{kl}&L^2_{kl}  
\\
\xi_k& 12& \widetilde{g}& g& -12& 0& g& \widetilde{g}& 12& -12& 0& 0& 0&0& 
P^4_{lk}& R^3_{lk}& L^2_{lk}&X^1_{kl} 
\end{array}
$$}

The submatrices are defined as follows:
$$
\begin{array}{lll}
P^1_{kl} = \R(e^{-\frac{4 i \pi}{25}} \z^{2(k+l)}) \ , &
P^2_{kl} = \R(e^{-\frac{2 i \pi}{25}} \z^{k+2l}) \ , &
P^3_{kl} = \R(e^{-\frac{8 i \pi}{25}} \z^{2l-k}) \ ,  
\\
P^4_{kl} = \R(e^{-\frac{6 i \pi}{25}} \z^{2(l-k)}) \ , &
R^1_{kl} = \R(e^{-\frac{i \pi}{25}} \z^{k+l}) \ , &
R^2_{kl} = \R(e^{-\frac{4 i \pi}{25}} \z^{l-k}) \ , 
\\
R^3_{kl} = \R(e^{-\frac{3 i \pi}{25}} \z^{l-2k}) \ ,&
L^1_{kl} = \R(e^{\frac{16 i \pi}{25}} \z^{k+l}) \ ,&
L^2_{kl} = \R(e^{\frac{12 i \pi}{25}} \z^{2k+l}) \ , 
\\
X^1_{kl} = \R(e^{\frac{9 i \pi}{25}} \z^{2(k+l)}) \ ,& &
\end{array}
$$
with indices $i,j=0,1,2$, $k,l=0,1,2,3$, 
and $g=6(1+\sqrt{5})$, $\tilde{g}=6(1-\sqrt{5})$.


\section{Superconformal $\TT$ and $\OO$ models}
\label{stoi-ch}
\subsection{Super-Tetrahedron}

In this Appendix we display the characters and the $S$ matrices for the 
superconformal \TT and \OO models. The characters are expressed in terms of 
the $\Th$ functions (\ref{th-def}) and of the Ising characters, 
$o$, $v$ and $s$ in Eq.(\ref{is-char}).

Untwisted NS sector $(i=0,1,2)$:
\be
\begin{array}{llll}
\chi_i &= \frac{1}{3}ooo
+ \frac{1}{6 \h} \left ( \w^i \Th\!\left[^{\ 0}_{1/3} \right]
+ \ov{\w}^i \Th\!\left[^{\ 0}_{2/3} \right] \right )(o+v)
& & 
\\
&\ \ + \frac{1}{6 \h} \left ( \w^i \Th\!\left[^{\ 0}_{5/6} \right]
+ \ov{\w}^i \Th\!\left[^{\ 0}_{1/6} \right] \right )(o-v)
\ ,& &h =0,2,2 \  , 
\\
\chi_- &= ovv \ ,& &h = 1 \ ,\\
\xi_i &=  \frac{1}{3}vvv
+  \frac{1}{6 \h} \left ( \w^i \Th\!\left[^{\ 0}_{1/3} \right]
+ \ov{\w}^i \Th\!\left[^{\ 0}_{2/3} \right] \right )(o+v)
& &
\\
& \ \ - \frac{1}{6 \h} \left ( \w^i \Th\!\left[^{\ 0}_{5/6} \right]
+ \ov{\w}^i \Th\!\left[^{\ 0}_{1/6} \right] \right )(o-v) 
\ ,& &h = \frac{3}{2}, \frac{5}{2}, \frac{5}{2} \ , 
\\
\xi_- &= voo  \ ,& &h = \frac{1}{2} \ .
\end{array}
\label{stet1}
\ee
Untwisted R sector $(i=0,1,2)$:
\be
\begin{array}{llll}
\r_i &= \frac{1}{3}sss
-  \frac{1}{3 \h} \left ( \w^i \Th\!\left[^{1/2}_{1/3} \right]
+ \ov{\w}^i \Th\!\left[^{1/2}_{2/3} \right] \right )s
\ ,& &h = \frac{19}{16}, \frac{19}{16}, \frac{3}{16} \ .  \\
\end{array}
\label{stet2}\ee
$\mathbb{Z}_2$-twisted NS sector: $\{ \l_s | \l =\s, \t \}$,
\be
\begin{array}{llll}
\s_s &= oss \ ,& &h = \frac{1}{8} \ , \\
\t_s &= vss \ ,& &h = \frac{5}{8} \ . \\
\end{array}
\label{stet3}\ee
$\mathbb{Z}_2$-twisted R sector: $\{ \s_I,\t_I |I=o,v \}$, 
\be
\begin{array}{llll}
\s_o &= soo \ ,& &h =  \frac{1}{16} \ , \\
\s_v &= osv \ ,& &h = \frac{9}{16} \  , \\
\t_o &= vso \ ,& &h = \frac{9}{16} \  , \\
\t_v &= svv \ ,& &h = \frac{17}{16} \  . \\
\end{array}
\label{stet4}
\ee
$\mathbb{Z}_3$-twisted NS sector ($ i = 0, 1, 2$):
\be
\begin{array}{llll}
\w^{\pm}_i  &= \frac{1}{6 \h} \left ( \Th\!\left[^{1/3}_{\ 0} \right] 
+ \w^i \Th\!\left[^{1/3}_{1/3} \right]
+ \ov{\w}^i \Th\!\left[^{1/3}_{2/3} \right] \right ) (o+v)
& & \\
&\ \pm \frac{1}{6 \h} \left ( \Th\!\left[^{1/3}_{1/2} \right] 
+\w^i \Th\!\left[^{1/3}_{5/6} \right]
+\ov{\w}^i \Th\!\left[^{1/3}_{1/6} \right] \right )(o-v)\ ,
& &  h = \frac{1}{18}, \frac{13}{18}, \frac{25}{18}, 
\frac{5}{9}, \frac{2}{9}, \frac{8}{9} \  , 
\\
\pi^{\pm}_i &= \frac{1}{6 \h} \left ( \Th\!\left[^{2/3}_{\ 0} \right] 
+ \w^i \Th\!\left[^{2/3}_{1/3} \right]
+ \ov{\w}^i \Th\!\left[^{2/3}_{2/3} \right] \right )(o+v)
&& 
\\
&\ \pm \frac{1}{6 \h} \left ( \Th\!\left[^{2/3}_{1/2} \right] 
+ \w^i \Th\!\left[^{2/3}_{5/6} \right]
+ \ov{\w}^i  \Th\!\left[^{2/3}_{1/6} \right] \right )(o-v) \ , 
& & h= \frac{2}{9}, \frac{8}{9}, \frac{5}{9}, 
\frac{13}{18}, \frac{25}{18}, \frac{1}{18} \ . 
\end{array} 
\label{stet5}
\ee
$\mathbb{Z}_3$-twisted R sector ($ i = 0, 1, 2$):
\be
\begin{array}{llll}
\l_i &=\frac{1}{3 \h} \left ( \Th\!\left[^{5/6}_{\ 0} \right] 
+\w^i \Th\!\left[^{5/6}_{1/3} \right]
+\ov{\w}^i \Th\!\left[^{5/6}_{2/3} \right] \right )s \ ,& &
h =\frac{59}{144}, \frac{11}{144}, \frac{107}{144} \  , \\
\psi_i   &= \frac{1}{3 \h} \left ( \Th\!\left[^{1/6}_{\ 0} \right] 
+\w^i \Th\!\left[^{1/6}_{1/3} \right]
+\ov{\w}^i \Th\!\left[^{1/6}_{2/3} \right] \right )s  \ ,& &
h =\frac{11}{144}, \frac{107}{144}, \frac{59}{144} \  . 
\end{array}
\label{stet6}
\ee

The $S$ matrix in this basis is reported in Table \ref{stetra-s}.

\begin{landscape}
\begin{table}
{\scriptsize
$$
\begin{array}{c|cccccccccccc}
& \c_j & \c_- & \x_j & \x_- & \r_j & \m_s & \s_J & \t_J & 
\w^\pm_j & \p^\pm_j & \l_j& \psi_j
\\ \hline
\c_i  & 1 & 3 & 1 & 3 & 2\sqrt{2} & 6 & 3\sqrt{2} & 3\sqrt{2} & 
4\w^i & 4 \ov{\w}^i & 4\sqrt{2}\w^i & 4\sqrt{2} \ov{\w}^i
\\
\c_-  & 3 & 9 & 3 & 9 & 6\sqrt{2} & -6 & -3\sqrt{2} & -3\sqrt{2} & 
0 & 0 & 0 & 0
\\ 
\x_i  & 1 & 3 & 1 & 3 & -2\sqrt{2} & 6 & -3\sqrt{2} & -3\sqrt{2} & 
4\w^i & 4 \ov{\w}^i & -4\sqrt{2}\w^i & -4\sqrt{2} \ov{\w}^i
\\ 
\x_-  & 3 & 9 & 3 & 9 & -6\sqrt{2} & -6 & 3\sqrt{2}& 3\sqrt{2}& 
0 & 0 & 0 & 0
\\
\r_i & 2\sqrt{2} & 6\sqrt{2} & -2\sqrt{2} & -6\sqrt{2} & 
0 & 0 & 0 & 0 & 
\pm 4\sqrt{2}\w^{i+1} & \mp 4\sqrt{2} \ov{\w}^{i+i} & 0 & 0
\\
\l_s & 6 & -6 & 6 & -6 & 0 & 0 & 
-6\sqrt{2}\e_{\l J} & 6\sqrt{2}\e_{\l J} & 0 & 0 & 0 & 0
\\
\s_I & 3\sqrt{2} & -3\sqrt{2} & -3\sqrt{2} & 3\sqrt{2} & 
0 & -6\sqrt{2}\e_{I \m} & 12 \d_{IJ} & 12(1-\d_{IJ}) &
0 & 0 & 0 & 0
\\
\t_I & 3\sqrt{2} & -3\sqrt{2} & -3\sqrt{2} & 3\sqrt{2} & 
0 & 6\sqrt{2}\e_{I \m} &  12(1-\d_{IJ}) & 12 \d_{IJ} &
0 & 0 & 0 & 0
\\
\w^\pm_i & 4\w^j & 0 & 4\w^j & 0 & \pm 4\sqrt{2}\w^{j+1} & 0 & 0 & 0 &
4\ov{\a} \w^{i+j} & 4\ov{\a}^2 \ov{\w}^{i+j} &
\pm 4\sqrt{2}\b \w^{i+j-1} & \pm 4\sqrt{2}\ov{\b} \ov{\w}^{i+j}
\\
\p^\pm_i & 4\ov{\w}^j & 0 & 4\w^j & 0 & \mp 4\sqrt{2}\ov{\w}^{j+1} & 0&0&0& 
4\ov{\a}^2 \ov{\w}^{i+j} & 4\ov{\a}^4 \w^{i+j}& 
\pm 4\sqrt{2}\b^8 \ov{\w}^{i+j}  & \pm 4\sqrt{2} \ov{\b}^2 \w^{i+j} 
\\
\l_i   & 4\sqrt{2}\w^j& 0 & -4\sqrt{2}\w^j & 0 & 0 & 0 & 0 & 0 &
\pm 4\sqrt{2}\b \w^{i+j-1}  & \pm 4\sqrt{2}\b^8 \ov{\w}^{i+j}& 
0 &0
\\
\psi_i & 4\sqrt{2}\w^j& 0 & -4\sqrt{2}\ov{\w}^j & 0 & 0 & 0 & 0 & 0 &
\pm 4\sqrt{2}\ov{\b} \ov{\w}^{i+j} & \pm 4\sqrt{2} \ov{\b}^2 \w^{i+j} &
0&0
\end{array}
$$}

\caption{
Super-tetrahedron $S$ matrix $(\times 24)$: $\e_{ij}=(-1)^{i+j}$, 
$\w=\exp(2i\pi/3)$, $\a=\exp(2i\pi/9)$ and $\b=\exp(i\pi/9)$,
with indices $i,j=0,1,2$; $\l,\m=\s,\t$ (resp. $0,1$) and
$I,J=o,v$ (resp. $0,1$).}
\label{stetra-s}
\end{table}
\end{landscape}


\subsection{Super-Octahedron}
Untwisted NS sector:
\be
\begin{array}{llll}
u_{\pm} &=  \frac{1}{6}ooo \pm \frac{1}{2}o(2 \t)o 
+ \frac{1}{6 \h} \Th\!\left[^{\ 0}_{1/3}\right](o+v)
+ \frac{1}{6 \h} \Th\!\left[^{\ 0}_{5/6}\right](o-v) \ ,&&h = 0, 5 \ , \\
u       &=   \frac{1}{3}ooo 
- \frac{1}{6 \h} \Th\!\left[^{\ 0}_{1/3}\right](o+v)
- \frac{1}{6 \h} \Th\!\left[^{\ 0}_{5/6}\right](o-v) \ ,&&h = 2 \ , \\
j_{\pm} &= \frac{1}{2}vvo \pm \frac{1}{2}v(2\t)o  \ ,& &h =1,2 \ , \\
v_{\pm} &= \frac{1}{6}vvv \pm \frac{1}{2}v(2 \t)v 
+ \frac{1}{6 \h} \Th\!\left[^{\ 0}_{1/3}\right](o+v)
- \frac{1}{6 \h} \Th\!\left[^{\ 0}_{5/6}\right](o-v)  \ ,
& &h = \frac{3}{2}, \frac{9}{2} \ , \\
v       &= \frac{1}{3}vvv - \frac{1}{6 \h} \Th\!\left[^{\ 0}_{1/3}\right](o+v)
+ \frac{1}{6 \h} \Th\!\left[^{\ 0}_{5/6}\right](o-v)  \ ,&&h = \frac{5}{2} \ , \\
h_{\pm} &=  \frac{1}{2}oov \pm \frac{1}{2}o(2\t)v  \ ,&&
h =  \frac{1}{2}, \frac{5}{2} \ . \\
\end{array}
\label{soct1}
\ee
Untwisted R sector:
\be
\begin{array}{llll}
\r_{\pm} &= \frac{1}{6}sss \pm \frac{1}{2}s(2 \t)s
- \frac{\bar{\w}}{3 \h} \Th\!\left[^{1/2}_{1/3}\right]s  \ ,& &
h =\frac{3}{16}, \frac{51}{16} \ , \\
\r       &=  \frac{1}{3}sss+
\frac{\bar{\w}}{3 \h} \Th\!\left[^{1/2}_{1/3}\right]s \ ,&&
h =\frac{19}{16} . \\
\end{array}
\label{soct2}
\ee
$\mathbb{Z}_3$-twisted NS sector ($ i = 0, 1, 2$):
\be
\begin{array}{llll}
\w^{\pm}_i  &= \frac{1}{6 \h} \left ( \Th\!\left[^{1/3}_{\ 0} \right] 
+ \w^i \Th\!\left[^{1/3}_{1/3} \right]
+ \ov{\w}^i \Th\!\left[^{1/3}_{2/3} \right] \right )(o+v)
& &
\\
& \ \pm \frac{1}{6 \h} \left ( \Th\!\left[^{1/3}_{1/2} \right] 
+ \w^i \Th\!\left[^{1/3}_{5/6} \right]
+ \ov{\w}^i \Th\!\left[^{1/3}_{1/6} \right] \right )(o-v)\ , 
& & h = \frac{1}{18}, \frac{13}{18}, \frac{25}{18}, 
\frac{5}{9}, \frac{2}{9}, \frac{8}{9} \ .
\end{array}
\label{soct5}
\ee
$\mathbb{Z}_3$-twisted R sector ($ i = 0, 1, 2$):
\be
\begin{array}{llll}
\l_i   &= \frac{1}{3 \h} \left ( \Th\!\left[^{5/6}_{\ 0} \right] 
+\w^i \Th\!\left[^{5/6}_{1/3} \right]
+\ov{\w}^i \Th\!\left[^{5/6}_{2/3} \right] \right )s   \ ,& &
h =\frac{59}{144}, \frac{11}{144}, \frac{107}{144} \  . \\
\end{array}
\label{soct6}
\ee
$\mathbb{Z}_4$-twisted NS sector ($ k = 0, 1, 2, 3$): 
\be
\begin{array}{llll}
\s_{s\pm}  &= \frac{1}{2}(sso \pm s(2\t)o) \ ,& 
&h = \frac{1}{8}, \frac{9}{8} \ , \\
\t_{s\pm}  &= \frac{1}{2}(ssv \pm s(2\t)v) \ ,& 
&h =\frac{5}{8}, \frac{13}{8} \ , \\
\a_k        &= \frac{1}{4 \h} \left (  
\Th\!\left[^{1/4}_{\ 0} \right]+i^k\Th\!\left[^{1/4}_{1/4} \right]
+(-1)^k\Th\!\left[^{1/4}_{1/2} \right]+(-i)^k\Th\!\left[^{1/4}_{3/4} \right]   
\right )o \ ,&&
h = \frac{1}{32}, \frac{9}{32}, \frac{49}{32}, \frac{25}{32} \ , \\
\b_k        &= \frac{1}{4 \h} \left (  
\Th\!\left[^{1/4}_{\ 0} \right]+i^k\Th\!\left[^{1/4}_{1/4} \right]
+(-1)^k\Th\!\left[^{1/4}_{1/2} \right]+(-i)^k\Th\!\left[^{1/4}_{3/4} \right]   
\right )v \ ,&
&h =\frac{17}{32}, \frac{25}{32}, \frac{65}{32}, \frac{41}{32}  \ . \\
\end{array}
\label{soct7}
\ee
$\mathbb{Z}_4$-twisted R sector ($I=o,v$; $ k = 0, 1, 2, 3$):
\be
\begin{array}{llll}
\s_{o\pm}  &= \frac{1}{2}(soo \pm o(2\t)s) \ ,& &
h = \frac{1}{16}, \frac{33}{16} \ , \\
\t_{v\pm}  &= \frac{1}{2}(svv \pm v(2\t)s) \ ,& &
h =\frac{17}{16}, \frac{33}{16} \ , \\
osv         &= osv \ ,& &h = \frac{9}{16} \ , \\
\g_k        &= \frac{1}{4 \h} \left (  
\Th\!\left[^{1/4}_{\ 0} \right]+i^k\Th\!\left[^{1/4}_{1/4} \right]
+(-1)^k\Th\!\left[^{1/4}_{1/2} \right]+(-i)^k\Th\!\left[^{1/4}_{3/4} \right]   
\right )s \ ,& &
h =\frac{3}{32}, \frac{11}{32}, \frac{51}{32}, \frac{27}{32}  \ . \\
\end{array}
\label{soct8}
\ee
$\mathbb{Z}_2$-twisted NS sector ($n=0,1$):
\be
\begin{array}{llll}
\m_{0 s} &= oss \ ,& &h = \frac{1}{8} \ , \\
\m_{1 s} &= vss \ ,& &h = \frac{5}{8} \ . \\
\end{array}
\label{soct9}\ee
$\mathbb{Z}_2$-twisted R sector ($I=o,v$, $n=0,1$):
\be
\begin{array}{llll}
\m_{0 o} &= soo \ ,& &h =  \frac{1}{16} \ , \\
\m_{0 v} &= osv \ ,& &h = \frac{9}{16} \  , \\
\m_{1 o} &= vso \ ,& &h = \frac{9}{16} \  , \\
\m_{1 v} &= svv \ ,& &h = \frac{17}{16} \  . \\
\end{array}
\label{soct10}
\ee

The $S$ matrix in this basis is reported in Table \ref{socta-s}.

\begin{landscape}
\begin{table}

{\tiny
$$
\begin{array}{c|ccccccccccccccccccc}
       & u_\pm & u & j_\pm & v_\pm & v & h_\pm & 
\r_\pm & \r    & \s_{s\pm} &\t_{s\pm} & \s_{I\pm} & osv &
\w^\b_j  & \l_j & \a_l & \b_l & \g_l & \m_{ms} & \mu_{mJ} 
\\ \hline
u_\pm & 1 & 2 & 3 & 1 & 2& 3& 2\sqrt{2}&4\sqrt{2}&6&6 & 3\sqrt{2}&6\sqrt{2} & 
8 & 8\sqrt{2} & \pm 6 & 
\pm 6 & \pm 6\sqrt{2} & \pm 12 &  \pm 6\sqrt{2}
\\
u & 2 & 4 & 6 & 2 & 4& 6& 4\sqrt{2}&8\sqrt{2}&12 &12&6\sqrt{2} & 12\sqrt{2} & 
-8 & -8\sqrt{2} & 0 & 
0 & 0 & 0 & 0
\\
j_\pm & 3 & 6 & 9 &3&6&9& 6\sqrt{2}&12\sqrt{2}&-6&-6& -3\sqrt{2} &-6\sqrt{2} & 
0 & 0 & \pm 6 & 
\pm 6 & \pm 6\sqrt{2}& \mp 12 & \mp 6\sqrt{2}
\\
v_\pm & 1 & 2 & 3 & 1 & 2& 3&-2\sqrt{2}&-4\sqrt{2}&6&6&-3\sqrt{2}&-6\sqrt{2} & 
8 & -8\sqrt{2} & \pm 6 & 
\pm 6 & \mp 6\sqrt{2} & \pm 12 &  \mp 6\sqrt{2}
\\
v & 2 & 4 & 6 & 2 & 4& 6&-4\sqrt{2}&-8\sqrt{2}&12 &12&-6\sqrt{2}&-12\sqrt{2} & 
-8 & 8\sqrt{2} & 0 & 
0 & 0 & 0 & 0
\\
h_\pm & 3 & 6 & 9 &3&6&9&-6\sqrt{2}&-12\sqrt{2}&-6&-6& 3\sqrt{2} & 6\sqrt{2} & 
0 & 0 & \pm 6 & 
\pm 6 & \mp 6\sqrt{2}& \mp 12 & \pm 6\sqrt{2}
\\
\r_\pm & 2\sqrt{2} & 4\sqrt{2} & 6\sqrt{2}&-2\sqrt{2}&-4\sqrt{2}&-6\sqrt{2} &
0 & 0& 0 & 0 & 0 & 0 & 
\e_\b 8\sqrt{2} & 0 &\pm \e_l 12 & 
\mp \e_l 12 &0 & 0 & 0
\\
\r & 4\sqrt{2} & 8\sqrt{2} & 12\sqrt{2}&-4\sqrt{2}&-8\sqrt{2}&-12\sqrt{2} &
0 & 0 & 0& 0 & 0 & 0 & 
-\e_\b 8\sqrt{2} & 0 &0& 
0 &0 & 0 & 0
\\
\s_{s\pm} &6&12&-6& 6 & 12& -6 &0&0&0&0 & 6\sqrt{2} & -12\sqrt{2} & 
0 & 0 & \pm \e_l 6\sqrt{2}&
\pm \e_l 6\sqrt{2} & \pm \e_l 12 & 0 & 0
\\
\t_{s\pm} &6&12&-6& 6 & 12& -6 &0&0&0&0 & -6\sqrt{2} & 12\sqrt{2} & 
0 & 0& \pm \e_l 6\sqrt{2}&
\pm \e_l 6\sqrt{2} & \mp \e_l 12 & 0 & 0
\\
\s_{I\pm} &3\sqrt{2}&6\sqrt{2}&-3\sqrt{2}&-3\sqrt{2}&-6\sqrt{2} &3\sqrt{2}&
0&0& 6\sqrt{2}& -6\sqrt{2}& \e_{IJ} 12 & 
0 & 0 & 0 & \pm 6\sqrt{2}&
\mp 6\sqrt{2} & 0 & 0 & \pm \e_{IJ}
\\
osv &6\sqrt{2}&12\sqrt{2}&-6\sqrt{2}&-6\sqrt{2}&-12\sqrt{2} &6\sqrt{2}&
0&0& -12\sqrt{2}& 12\sqrt{2}& 0 & 
0 & 0 & 0 & 0 &
0 & 0 & 0 & 0
\\
\w^\a_i &8 &-8 & 0&8 &-8 &0 &
\e_\a 8\sqrt{2} &-\e_\a 8\sqrt{2} &0 &0 & 0 & 0 &
w_{ij} &\e_\a l_{ij}  & 0 & 
0 & 0 & 0 & 0
\\
\l_i &8 \sqrt{2}&-8 \sqrt{2}& 0&-8\sqrt{2}&8\sqrt{2} &0 &
0 & 0&0 &0 & 0 & 0 &
\e_\b l_{ij}  & 0 & 0&
0 & 0 & 0 & 0
\\
\a_k &\pm 6 &0 & \pm 6& \pm 6 &0 &\pm 6 &\pm \e_k 12 & 0& 
\pm \e_k 6\sqrt{2}&\pm \e_k 6\sqrt{2}& \pm 6\sqrt{2}& 0 &
0 & 0 & a_{kl}&
a_{kl} & \sqrt{2}a_{kl} & 0 & 0
\\
\b_k &\pm 6 &0 & \pm 6& \pm 6 &0 &\pm 6 &\mp \e_k 12 & 0& 
\pm \e_k 6\sqrt{2}&\pm \e_k 6\sqrt{2}& \mp 6\sqrt{2}& 0 &
0 & 0 & a_{kl}&
a_{kl} & -\sqrt{2}a_{kl} & 0 & 0
\\
\g_k &\pm 6\sqrt{2} &0 & \pm 6\sqrt{2}& \mp 6\sqrt{2} &0 & 
\mp 6\sqrt{2} &0&0 &\pm \e_k 12 &\mp \e_k 12& 0&0 & 
0 & 0 & \sqrt{2} a_{kl}&
-\sqrt{2}a_{kl} & 0 & 0 & 0
\\
\m_{n s} &\pm 12 &0 & \mp 12& \mp 12 &0 & 
\mp 12 &0&0 &0 &0 & 0&0 & 
0 & 0 & 0&
0 & 0 & 0 & \e_{nmJ} 12\sqrt{2}
\\
\m_{n I} &\pm 6\sqrt{2} &0 & \mp 6\sqrt{2}& \mp 6\sqrt{2} &0 & 
\pm 6\sqrt{2} &0&0 &0 &0 & \pm \e_{IJ}&0 & 
0 & 0 & 0& 0 & 0 & \e_{nmI} 12\sqrt{2} & \e_{nm} 12
\end{array}
$$}
\caption{Super-octahedron $S$ matrix  $(\times 48)$.
The sub-matrices are defined as follows:
$w_{ij} = 16 \R\left(e^{-2i\pi/9} \w^{i+j} \right)$ ,
$l_{ij} = 16 \sqrt{2} \R\left(e^{-5i\pi/9} \w^{i+j} \right)$ , and
$a_{kl} = 12 \R \left( e^{-i\pi/8} i^{k+l} \right)$,
with indices $i,j=0,1,2$; $k,l=0,1,2,3$, and $n,m=0,1$;
the $\e_{ijk...}$ are signs defined according to 
$\e_{ijk...}=(-1)^{i+j+k+\dots}$.
The indices $\a,\b=+,-$, in $\w^\a_i,w^\b_j$ and 
$I,J=o,v$, in $\s_{I\pm}$ and $\m_{nI}$ should be 
considered as taking the values $0,1$, respectively.
}
\label{socta-s}
\end{table}
\end{landscape}



\def\NPB#1#2#3{{\it Nucl.~Phys.} {\bf{B#1}} (#2) #3}
\def\CMP#1#2#3{{\it Commun.~Math.~Phys.} {\bf{#1}} (#2) #3}
\def\CQG#1#2#3{{\it Class.~Quantum~Grav.} {\bf{#1}} (#2) #3}
\def\PLB#1#2#3{{\it Phys.~Lett.} {\bf{B#1}} (#2) #3}
\def\PRD#1#2#3{{\it Phys.~Rev.} {\bf{D#1}} (#2) #3}
\def\PRL#1#2#3{{\it Phys.~Rev.~Lett.} {\bf{#1}} (#2) #3}
\def\ZPC#1#2#3{{\it Z.~Phys.} {\bf C#1} (#2) #3}
\def\PTP#1#2#3{{\it Prog.~Theor.~Phys.} {\bf#1}  (#2) #3}
\def\MPLA#1#2#3{{\it Mod.~Phys.~Lett.} {\bf#1} (#2) #3}
\def\PR#1#2#3{{\it Phys.~Rep.} {\bf#1} (#2) #3}
\def\AP#1#2#3{{\it Ann.~Phys.} {\bf#1} (#2) #3}
\def\RMP#1#2#3{{\it Rev.~Mod.~Phys.} {\bf#1} (#2) #3}
\def\HPA#1#2#3{{\it Helv.~Phys.~Acta} {\bf#1} (#2) #3}
\def\JETPL#1#2#3{{\it JETP~Lett.} {\bf#1} (#2) #3}
\def\JHEP#1#2#3{{\it JHEP} {\bf#1} (#2) #3}
\def\TH#1{{\tt hep-th/#1}}

\end{document}

%% file: moduli.pstex_t
\begin{picture}(0,0)%
\includegraphics{moduli.pstex}%
\end{picture}%
\setlength{\unitlength}{2763sp}%
\begingroup\makeatletter\ifx\SetFigFont\undefined%
\gdef\SetFigFont#1#2#3#4#5{%
  \reset@font\fontsize{#1}{#2pt}%
  \fontfamily{#3}\fontseries{#4}\fontshape{#5}%
  \selectfont}%
\fi\endgroup%
\begin{picture}(6708,7266)(1726,-7594)
\put(2476,-5536){\makebox(0,0)[lb]{\smash{\SetFigFont{8}{9.6}{\familydefault}{\mddefault}{\updefault}{\large $\frac{1}{\sqrt{2}}$}}}}
\put(1726,-7411){\makebox(0,0)[lb]{\smash{\SetFigFont{8}{9.6}{\familydefault}{\mddefault}{\updefault}{\large $R_{sa}$}}}}
\put(1726,-5236){\makebox(0,0)[lb]{\smash{\SetFigFont{8}{9.6}{\familydefault}{\mddefault}{\updefault}{\large $\sqrt{2}$}}}}
\put(1876,-4036){\makebox(0,0)[lb]{\smash{\SetFigFont{8}{9.6}{\familydefault}{\mddefault}{\updefault}{\large $1$}}}}
\put(4351,-7411){\makebox(0,0)[lb]{\smash{\SetFigFont{8}{9.6}{\familydefault}{\mddefault}{\updefault}{\large $R_{so}$}}}}
\put(5926,-3661){\makebox(0,0)[lb]{\smash{\SetFigFont{8}{9.6}{\familydefault}{\mddefault}{\updefault}{\large $\frac{1}{\sqrt{2}}$}}}}
\put(5626,-3361){\makebox(0,0)[lb]{\smash{\SetFigFont{8}{9.6}{\familydefault}{\mddefault}{\updefault}{\large $1$}}}}
\put(5551,-661){\makebox(0,0)[lb]{\smash{\SetFigFont{8}{9.6}{\familydefault}{\mddefault}{\updefault}{\large $R_o$}}}}
\put(7951,-4336){\makebox(0,0)[lb]{\smash{\SetFigFont{8}{9.6}{\familydefault}{\mddefault}{\updefault}{\large $R_c$}}}}
\put(4276,-5611){\makebox(0,0)[lb]{\smash{\SetFigFont{8}{9.6}{\familydefault}{\mddefault}{\updefault}{\large $\sqrt{2}$}}}}
\put(6751,-5611){\makebox(0,0)[lb]{\smash{\SetFigFont{8}{9.6}{\familydefault}{\mddefault}{\updefault}{\large $R_{o'}$}}}}
\put(4351,-4336){\makebox(0,0)[lb]{\smash{\SetFigFont{8}{9.6}{\familydefault}{\mddefault}{\updefault}{\large $1$}}}}
\put(3376,-4336){\makebox(0,0)[lb]{\smash{\SetFigFont{8}{9.6}{\familydefault}{\mddefault}{\updefault}{\large $\frac{1}{\sqrt{2}}$}}}}
\put(5176,-4336){\makebox(0,0)[lb]{\smash{\SetFigFont{8}{9.6}{\familydefault}{\mddefault}{\updefault}{\large $\sqrt{2}$}}}}
\end{picture}